\pdfoutput=1

\documentclass[11pt,twoside,a4paper,cmspaper,final,collab]{cms-tdr}
\def\svnVersion{191425}\def\cmsCernNo{2013-050}\def\cmsCernDate{\today}\def\cmsMessage{Published in the European Physical Journal C as \href{http://dx.doi.org/10.1140/epjc/s10052-013-2469-8}{\doi{10.1140/epjc/s10052-013-2469-8.}}}

\begin{document}\cmsNoteHeader{HIG-12-034}

\hyphenation{had-ron-i-za-tion}
\hyphenation{cal-or-i-me-ter}
\hyphenation{de-vices}

\RCS$Revision: 189435 $
\RCS$HeadURL: svn+ssh://svn.cern.ch/reps/tdr2/papers/HIG-12-034/trunk/HIG-12-034.tex $
\RCS$Id: HIG-12-034.tex 189435 2013-06-08 18:44:16Z asavin $
\newlength\cmsFigWidth
\ifthenelse{\boolean{cms@external}}{\setlength\cmsFigWidth{0.49\textwidth}}{\setlength\cmsFigWidth{0.6\textwidth}}
\ifthenelse{\boolean{cms@external}}{\providecommand{\cmsLeft}{top\xspace}}{\providecommand{\cmsLeft}{left\xspace}}
\ifthenelse{\boolean{cms@external}}{\providecommand{\cmsRight}{bottom\xspace}}{\providecommand{\cmsRight}{right\xspace}}
\ifthenelse{\boolean{cms@external}}{\providecommand{\breakhere}{\linebreak[4]}}{\providecommand{\breakhere}{\relax}}
\providecommand{\MT}{\ensuremath{M_{\mathrm{T}}\xspace}}
\providecommand{\vecEtm}{\ensuremath{\vec{E}_\mathrm{T}^{\text{miss}}}\xspace}
\providecommand{\vecPtell}{\ensuremath{\vec{p}_\mathrm{T}^{\ell\ell}}\xspace}
\providecommand{\vecPtel}{\ensuremath{\vec{p}_\mathrm{T}^{\ell}}\xspace}
\providecommand{\re}{\cmsSymbolFace{e}}
\providecommand{\mjj}{\ensuremath{m_\mathrm{jj}}\xspace}
\providecommand{\mt}{\ensuremath{m_\mathrm{T}}\xspace}
\newcommand{\Zee}{\ensuremath{\cPZ\to\Pep\Pem}}
\newcommand{\mW}{\ensuremath{m_\PW}}
\newcommand{\mZZ}{\ensuremath{m_{\cPZ\cPZ}}}
\newcommand{\mll}{\ensuremath{m_{\ell \ell}}}
\newcommand{\tauh}{\ensuremath{\Pgt_\mathrm{h}}}
\newcommand{\taue}{\ensuremath{\Pgt_\Pe}}
\newcommand{\taumu}{\ensuremath{\Pgt_\Pgm}}
\newcommand{\taul}{\ensuremath{\Pgt_\ell}}
\newcommand{\Hmu}{\ensuremath{\PH\to\cPZ\cPZ^{(\ast)}\to 4\Pgm}}
\newcommand{\mmu}{\ensuremath{m_{4\Pgm}}}
\newcommand{\mH}{\ensuremath{m_{\PH}}}
\newcommand{\lnQ}{\ensuremath{\ln(Q)}}
\newcommand{\mlnQ}{\ensuremath{-2\,\ln(Q)}}
\newcommand{\LS}{\ensuremath{\mathcal{L}(S)}}
\newcommand{\LB}{\ensuremath{\mathcal{L}(B)}}
\newcommand{\LSB}{\ensuremath{\mathcal{L}(S+B)}}
\newcommand{\mZ}{\ensuremath{m_{\cPZ}}}
\newcommand{\Zbb}{\ensuremath{(\cPZ/\Pgg^*){\cPqb\cPaqb}}}
\newcommand{\ZZ}{\ensuremath{\cPZ\cPZ}}
\newcommand{\Zgam}{\ensuremath{\cPZ/\Pgg^*}}
\newcommand{\mumu}{\ensuremath{\Pgmp\Pgmm}}
\newcommand{\Wo}{\ensuremath{\PW}}
\newcommand{\Wp}{\ensuremath{\PWp}}
\newcommand{\Wm}{\ensuremath{\PWm}}
\newcommand{\Zo}{\ensuremath{\cPZ}}
\newcommand{\Ho}{\ensuremath{\PH}}
\newcommand{\KD}{\mathrm{KD} }
\newcommand{\X}{\mathrm{X} }
\newcommand{\Tau}{\ensuremath{\tau}}
\newcommand{\Lep}{\ensuremath{\ell}}
\newcommand{\ptlmax}{\ensuremath{p_{\mathrm{T}}^{\Lep,\mathrm{max}}}}
\newcommand{\GAMMA}{\ensuremath{\gamma}}
\newcommand{\Hi}{\PH}
\newcommand{\Mp}{\Pgmp}
\newcommand{\Mm}{\Pgmm}
\newcommand{\deletall}{\ensuremath{\Delta\eta_{\Lep\Lep}}}
\newcommand{\delphill}{\ensuremath{\Delta\phi_{\Lep\Lep}}}
\newcommand{\delphil}{\ensuremath{\Delta\phi_{\Lep}}}
\newcommand{\delphimetll}{\ensuremath{\Delta\phi_{\Lep\Lep,\MET}}}
\newcommand{\delphimetl}{\ensuremath{\Delta\phi_{\Lep,\MET}}}
\newcommand{\ptlmin}{\ensuremath{p_{\mathrm{T}}^{\Lep,\text{min}}}}
\newcommand{\wgamma}{\ensuremath{\Wo\GAMMA}}
\newcommand{\WW}{\ensuremath{\Wo\Wo}}
\newcommand{\WpWm}{\ensuremath{\Wo^+\Wo^-}}
\newcommand{\tw}{\ensuremath{\cPqt\Wo}}
\newcommand{\WZ}{\ensuremath{\Wo\Zo}}
\newcommand{\Wjets}{\PW+\text{jets}}
\newcommand{\Hww}{\Hi\to\WW}
\newcommand{\dytt}{\ensuremath{\cPZ/\GAMMA^* \to\tau^+\tau^-}}
\newcommand{\dyll}{\ensuremath{\cPZ/\GAMMA^*\mathrm{\to \ell^+\ell^-}}}
\newcommand{\dyee}{\ensuremath{\cPZ/\GAMMA^*\mathrm{\to e^+e^-}}}
\newcommand{\METpr}{\ensuremath{\mathrm{E}_{\text{T, projected}}^{\text{miss}}}}

\cmsNoteHeader{HIG-12-034} 
\title{Search for a standard-model-like Higgs boson with a mass in the range 145 to 1000\GeV at the LHC}

\date{\today}

\abstract{
A search for a standard-model-like Higgs boson in the $\PH \to \WW$ and $\PH \to \ZZ$ decay channels
is reported, for Higgs boson masses in the range $145 < \mH < 1000\GeV$. The search is based upon
proton-proton collision data samples corresponding to an integrated luminosity
of up to 5.1\fbinv at $\sqrt{s} = 7\TeV$
and up to 5.3\fbinv at $\sqrt{s} = 8\TeV$, recorded by the CMS experiment at the LHC.
The combined
upper limits at 95\% confidence level on products of the cross section and branching
fractions
exclude a standard-model-like Higgs boson in the range
$145 < \mH <  710\GeV$, thus extending the
mass region excluded by CMS from
127--600\GeV up to 710\GeV.
}

\hypersetup{%
pdfauthor={CMS Collaboration},%
pdftitle={Search for a standard-model-like Higgs boson with a mass in the range 145 to 1000 GeV at the LHC}, %
pdfsubject={CMS},%
pdfkeywords={CMS, physics, Higgs, multilepton, diboson}}

\maketitle 

\section{Introduction}
\label{sec:introduction}

The standard model (SM) of electroweak interactions~\cite{StandardModel67_1,StandardModel67_2,StandardModel67_3}
relies on the existence of the Higgs boson, $\PH$, a scalar particle associated with the field responsible for
spontaneous electroweak symmetry breaking~\cite{Englert:1964et,Higgs:1964ia,Higgs:1964pj,Guralnik:1964eu,Higgs:1966ev,Kibble:1967sv}.
The mass of the boson, $\mH$, is not predicted by the theory. Searches for the SM Higgs boson at LEP and the Tevatron
excluded at 95\% confidence level (CL) masses lower than 114.4\GeV~\cite{Barate:2003sz} and the mass range 162--166\GeV~\cite{TEVHIGGS_2010}, respectively. Previous direct searches at the Large Hadron Collider (LHC)~\cite{Evans:2008zzb} were based on data from proton-proton (\Pp\Pp) collisions corresponding to
an integrated luminosity of up to 5\fbinv, collected at a center-of-mass energy $\sqrt{s}=7\TeV$.
Using the 7\TeV data set the Compact Muon Solenoid (CMS) experiment has excluded at 95\% CL masses from 127 to 600\GeV~\cite{Chatrchyan:2012tx}.
In 2012, the LHC $\Pp\Pp$ center-of-mass energy was increased to $\sqrt{s}=8\TeV$, and an additional integrated luminosity
of more than 5\fbinv was recorded by the end of June.
Searches based on these data in the mass range 110--145\GeV led to the
observation of a new boson with a mass of approximately
$125\GeV$~\cite{ATLASobservation125,CMSobservation125,CMSlongpaper}. 
Using this data set the ATLAS experiment
excluded at 95\% CL the mass ranges
111--122 and 131--559\GeV~\cite{ATLASobservation125}.
By the end of 2012 the amount of collected integrated luminosity at 8\TeV 
reached almost 20\fbinv. We intend to report findings from the
entire data set in a future publication. However, given the heightened interest 
following the recent discovery of the 125\GeV boson, and the fact that
the analysis of the full data taken in 2011--2012 will take time,
we present here a search for the SM-like Higgs boson 
up to 1\TeV with the same
data set that was used in 
Refs.~\cite{CMSobservation125,CMSlongpaper}.

The observation of a Higgs boson with a mass of 125\GeV is
consistent with the theoretical constraint coming from the unitarization
of diboson scattering at high
energies~\cite{Dicus:1992vj,Veltman:1976rt,Lee:1977eg,Lee:1977yc,Passarino:1990hk,Chanowitz:1985hj,Duncan:1985vj,Dicus:1986jg,Bagger:1995mk,Ballestrero:2009vw}.
However, there is still a
possibility that the newly discovered particle
has no connection to the electroweak symmetry breaking mechanism~\cite{Low:2011gn,Low:2012rj}.
In addition, several popular scenarios, such as general two-Higgs-doublet models (for a review see~\cite{Branco:2011iw,craig}) or models
in which the SM Higgs boson mixes with a heavy electroweak singlet~\cite{Patt:2006fw}, predict the existence
of additional resonances at high mass, with couplings similar to the SM Higgs boson. In any such models,
issues related to the width of the resonance and its interference with non-resonant $\WW$ and $\ZZ$ backgrounds must be understood.
This paper reports a search for a SM-like Higgs boson at
high mass, assuming the properties predicted by the SM. The $\PH \to \WW$ and $\PH \to \ZZ$ decay channels are used as
benchmarks for cross section and production mechanism in the mass range $145 < \mH < 1000$\GeV. This approach allows for a
self-consistent and coherent presentation of the results at high mass.

For a Higgs boson decaying to two $\PW$ bosons, the
fully leptonic ($\PH \to \WW \to \ell\nu\ell\nu$) and semileptonic ($\PH \to \WW \to \ell \nu
\cPq\cPq$) final states are considered in this analysis. For a Higgs boson decaying into two $\cPZ$
bosons, final states containing four leptons ($\PH \to \ZZ \to 2\ell 2\ell'$), two leptons and two
jets ($\PH \to \ZZ \to 2\ell 2\cPq$), and two leptons and two neutrinos
($\PH \to \ZZ \to 2\ell 2\nu$), are considered, where $\ell = \Pe$ or $\Pgm$ and $\ell' = \Pe$, $\Pgm$,
or $\Pgt$. The analyses use $\Pp\Pp$ collision data samples recorded by the CMS detector, corresponding to integrated
luminosities of up to 5.1\fbinv at $\sqrt{s} = 7\TeV$ and up to 5.3\fbinv at
$\sqrt{s} = 8\TeV$.

\section{ The CMS detector and simulations}
\label{sec:cms}

A full description of the CMS apparatus is
available elsewhere~\cite{Chatrchyan:2008zzk}.
The CMS experiment uses a right-handed coordinate system, with the origin at the nominal interaction point, the $x$ axis
pointing to the center of the LHC ring, the $y$ axis pointing up (perpendicular to the plane of the LHC ring),
and the $z$ axis along the counterclockwise-beam direction. The polar angle $\theta$ is measured from the
positive $z$ axis, and the azimuthal angle $\phi$ is measured in the $x$--$y$ plane. All angles in this paper are presented in radians. The pseudorapidity
is defined as $\eta=-\ln[\tan{(\theta/2)}]$.

\hyphenation{Cher-en-kov POW-HEG pa-ram-e-trized}
{\tolerance=800 The central feature of the CMS apparatus is a superconducting solenoid of 6\unit{m} internal diameter, which
provides a magnetic field of 3.8\unit{T}. Within the field volume are a silicon pixel and strip tracker,
a lead tungstate crystal electromagnetic calorimeter (ECAL), and a \breakhere{}brass/scintillator hadron calorimeter.
A quartz-fiber Cherenkov calorimeter extends the coverage to $\abs{\eta} < 5.0$. Muons are measured in
gas-ionization detectors embedded in the steel flux return yoke. The first level of the CMS trigger system, composed of custom hardware processors, is designed to select the most interesting events in less than $3\mus$, using information from the calorimeters and muon detectors. The high level trigger processor farm decreases the event rate from 100\unit{kHz} delivered by the first level trigger to a few hundred hertz, before data storage.\par}

Several Monte Carlo (MC) event generators are used to simulate the signal and background event samples.
The $\PH \to \WW$ and $\PH \to \ZZ$ signals are simulated using the next-to-leading order (NLO) package
{\POWHEG}~\cite{Alioli:2008gx,Nason:2004rx,Frixione:2007vw}. The Higgs boson signals from gluon fusion ($\Pg\Pg \to \PH$),
and vector-boson fusion (VBF, $\Pq\Pq \to \Pq\Pq \PH$), are generated with {\POWHEG} at NLO and a dedicated
program~\cite{Gao:2010qx} used for angular correlations. Samples of $\PW\PH$, $\cPZ\PH$, and $\ttbar\PH$ events are
generated using \PYTHIA 6.424~\cite{Sjostrand:2006za}.

At generator level, events are weighted according to the total cross section $\sigma(\Pp\Pp\to \PH)$,
which contains contributions from gluon fusion computed to next-to-next-to-leading order (NNLO) and next-to-next-to-leading-log
(NNLL)~\cite{Anastasiou:2008tj,deFlorian:2009hc,Baglio:2010ae,LHCHiggsCrossSectionWorkingGroup:2011ti,Djouadi:1991tka,Dawson:1990zj,Spira:1995rr,Harlander:2002wh,Anastasiou:2002yz,Ravindran:2003um,Catani:2003zt,Actis:2008ug}, and from
weak-boson fusion computed at NNLO~\cite{LHCHiggsCrossSectionWorkingGroup:2011ti,Ciccolini:2007jr,Ciccolini:2007ec,Figy:2003nv,Arnold:2008rz,Bolzoni:2010xr}.

The simulated \WW (\ZZ) invariant mass $m_{\WW}$ ($m_{\ZZ}$) lineshape is corrected to match the results
presented in Ref.~\cite{Passarino:2010qk,Goria:2011wa,Kauer:2012hd}, where the complex-pole scheme
for the Higgs boson propagator is used.
In the gluon fusion production channel,
the effects on the lineshape due to interference between Higgs boson signal and the  $\Pg\Pg \to \WW$ and $\Pg\Pg \to \ZZ$
backgrounds are included~\cite{Passarino:2012ri,Campbell:2011cu}. The theoretical uncertainties on the lineshape due to missing
higher-order corrections in the interference between background and signal are included in the total uncertainties, in addition to uncertainties
associated with electroweak corrections~\cite{Goria:2011wa,Passarino:2012ri}. Interference outside the Higgs boson mass
peak has sizable effects on the normalization for those final states where the Higgs boson invariant mass cannot be fully reconstructed. A correction is applied, taking into account the corresponding theoretical uncertainties,
in the $\WW \to \ell \nu \cPq\cPq$ final state~\cite{Passarino:2012ri,Campbell:2011cu}. In the $\WW \to \ell\nu\ell\nu$ and
$\ZZ \to 2\ell 2\nu$ final states, the effect of interference on the normalization, as computed in ~\cite{Campbell:2011cu,Kauer:2012ma}, is included with an associated uncertainty of 100\%.

The background contribution from $\cPq\cPaq \to \WW$ production is generated using the \MADGRAPH package~\cite{Alwall:2007st}, and the subdominant $\Pg\Pg \to \WW$ process is generated using \textsc{gg2ww}~\cite{ggww}.
The $\cPq\cPaq \to \ZZ$ production process is simulated at NLO
with \POWHEG, and the $\mathrm{gg} \to \ZZ$ process is simulated using
\textsc{gg2zz}~\cite{Binoth:2008pr}. Other diboson processes ($\PW\cPZ$, $\cPZ\gamma^{(*)}$, $\PW\gamma^{(*)}$) and $\cPZ+$jet are generated with \PYTHIA6.424 and \MADGRAPH. The $\ttbar$ and $\cPqt\PW$ events are generated at NLO with \POWHEG.
For all samples
\PYTHIA is used for parton showering, hadronization, and
 underlying event simulation.
For leading-order (LO) generators, the default set of parton distribution functions (PDF) used to produce these samples is CTEQ6L~\cite{CTEQ66}, while CT10~\cite{ct10} is used for NLO generators. The $\tau$-lepton decays are simulated with \TAUOLA~\cite{Jadach:1993hs}. The detector response is simulated using a detailed description of the CMS detector, based on the \GEANTfour package~\cite{GEANT}, with event reconstruction performed identically to that for recorded data.
The simulated samples include the effect of multiple $\Pp\Pp$ interactions per bunch crossing (pileup).
The \PYTHIA parameters for the underlying events and pileup interactions
are set to the Z2 ($\cPZ 2^*$) tune for the 7\,(8)\TeV data sample as described in Ref.~\cite{Chatrchyan:2011id}
with the pileup multiplicity distribution matching that seen in data.

\section{Event reconstruction}
\label{sec:reconstruction}

A complete reconstruction of the individual particles emerging from each collision event is obtained via a particle-flow (PF) technique~\cite{CMS-PAS-PFT-09-001, CMS-PAS-PFT-10-002}. This approach uses the information from all CMS sub-detectors to identify and reconstruct individual particles in the collision event, classifying them into mutually exclusive categories: charged hadrons, neutral hadrons, photons, electrons, and muons.

The electron reconstruction algorithm combines information from clusters of energy deposits in the ECAL with the
trajectory in the inner tracker~\cite{Baffioni:2006cd,CMS-PAS-EGM-10-004}. Trajectories in the tracker volume are reconstructed using a dedicated model of electron energy loss, and fitted with a Gaussian sum filter. Electron identification relies on a multivariate (MVA) technique that combines observables sensitive to the amount of bremsstrahlung
along the electron trajectory, the geometrical and momentum matching between the electron trajectory and
the associated clusters, and shower-shape observables.

The muon reconstruction algorithm combines information from the silicon tracker and the muon spectrometer. Muons are
selected from amongst the reconstructed muon-track candidates by applying requirements on the track components in
the muon system and on matched energy deposits in the calorimeters~\cite{CMS-PAS-PFT-10-003}.

The $\tau$-leptons are identified in both the leptonic decay modes, with an electron or muon as measurable decay
product, and in the hadronic mode (denoted $\tauh$). The PF particles are used to reconstruct $\tauh$ using the
``hadron-plus-strip'' (HPS) algorithm~\cite{Chatrchyan:2011xq}.

Jets are reconstructed from PF candidates by using the anti-\kt clustering algorithm
\cite{Cacciari:2008gp,fastjetmanual} with a distance parameter of 0.5. Jet energy corrections are applied to
account for the non-linear response of the calorimeters, and other
instrumental effects. These corrections are based on in-situ calibration using dijet and $\gamma/\cPZ+$jet data
samples~\cite{Chatrchyan:2011ds}. The median energy density due to pileup is evaluated in each event, and the
corresponding energy is subtracted from each jet~\cite{Cacciari:2008gn}. Jets are required to originate at the primary vertex, which is identified as the vertex with the highest summed $\PT^2$ of its associated tracks.
Jets displaced from the primary vertex in the transverse direction can be tagged as b jets~\cite{BTV}.

Charged leptons from $\PW$ and $\cPZ$ boson decays are typically expected to be isolated from other activity in the
event. The isolation of $\Pe$ or $\Pgm$ leptons is therefore ensured by applying requirements on the sum of the
transverse energies of all reconstructed particles, charged or neutral, within a cone of
$\Delta R = \sqrt{(\Delta \eta)^{2}
+ (\Delta \phi)^{2}} < 0.4$ around the lepton direction, after subtracting the average pileup energy
estimated using a ``jet area'' technique~\cite{Cacciari:2007fd} on an event-by-event basis.

The magnitude of the transverse momentum ($\PT$) is calculated as $\PT = \sqrt{\smash[b]{p_{x}}^2 + \smash[b]{p_{y}}^2}$.
The missing transverse energy vector $\vecEtm$ is defined as
the negative vector sum of the
transverse momenta of all reconstructed particles in the event,
with $\MET = |\vecEtm|$.

At trigger level, depending on the decay channel, events are required to have a pair of electrons or muons, or an electron and a muon,
one lepton with $\PT > 17\GeV$ and the other with $\PT > 8\GeV$, or a single electron (muon) with $\PT > 27\,(24)\GeV$.

The efficiencies for trigger selection, reconstruction, identification, and isolation of $\Pe$ and $\Pgm$
are measured from recorded data, using a ``tag-and-probe''~\cite{CMS:2011aa} technique based on an inclusive sample of
$\cPZ$-boson candidate events. These measurements are performed in several bins of $\PT^{\ell} $ and $ \abs{\eta^{\ell}} $.
The overall trigger efficiency
for events selected for this analysis
ranges from 96\% to 99\%.
The efficiency of the electron identification in the ECAL barrel (endcaps) varies from around
82\% (73\%) at $\PT^{\Pe} \simeq 10\GeV$
to 90\%\,(89\%) for
$\PT^{\Pe} \simeq 20\GeV$. It drops to about 85\% in the transition region, $1.44 < \abs{\eta^{\Pe}} <
1.57$, between the ECAL barrel and endcaps. Muons with $\pt>5\GeV$ are reconstructed and identified with efficiencies greater than
${\sim}98\%$ in the full $\abs{\eta^{\Pgm}} < 2.4$ range. The efficiency of the $\tauh$ identification is around 50\% for
$\pt^\Pgt > 20\GeV$~\cite{Chatrchyan:2011xq}.

\section{Data analysis}
\label{sec:analyses}

The results presented in this paper are obtained by combining Higgs boson searches exploiting different production and decay modes. A summary of these searches is given in Table~\ref{tab:channels}.
All final states are exclusive, with no overlap between channels.
The results of the searches in the mass range $\mH < 145\GeV$ are presented
in Ref.~\cite{CMSobservation125,CMSlongpaper}.
The presence of a signal in any one of the channels, at a certain value of the Higgs boson mass, is expected to manifest itself as an excess extending around that value for a range corresponding to the Higgs boson width convoluted with the experimental mass resolution. The Higgs boson width varies from few percents of $\mH$ at low masses through up to 50\% at $\mH=1\TeV$. The mass resolution for each decay mode is given in Table~\ref{tab:channels}.
It should be noted that the presence of the boson with $\mH = 125\GeV$
effectively constitutes an
additional background especially
in the $\PW\PW \to \ell\nu\ell\nu$ channel
up to approximately $\mH=200\GeV$,
because of the poor mass resolution of this analysis.
To take this effect explicitly into account a simulated SM Higgs boson signal with $\mH=125\GeV$ is considered
as background in this paper.

\begin{table*}[htbp]
\begin{center}
\small
\topcaption{Summary information on the analyses included in this paper. The column ``\PH production''
indicates the production mechanism targeted by an analysis; it does not imply 100\% purity. The main contribution in the untagged and inclusive categories is always gluon fusion.
The $(\mathrm{jj})_\mathrm{VBF}$ refers to dijet pair consistent with the VBF topology, and $(\mathrm{jj})_{\PW(\cPZ)}$ to a dijet pair with
an invariant mass consistent with coming from a \PW\ (\cPZ) dijet decay.
For the $\PW\PW \to \ell\nu\ell\nu$ and
$\cPZ\cPZ \to 2\ell 2\ell'$ channels the full possible mass range starts from  110\GeV, but in this paper
both analyses are restricted to the masses above 145\GeV. The
$\cPZ\cPZ \to 2\ell 2\cPq$ analysis uses only 7\TeV data. The notation ``((\Pe\Pe, \Pgm\Pgm), \Pe\Pgm) + (0 or 1 jets)'' indicates that the analysis is performed in two independent lepton categories (\Pe\Pe, \Pgm\Pgm) and (\Pe\Pgm), each category further subdivided in two subcategories with zero or one jets, thus giving a total of four independent channels. 
}
\label{tab:channels}
\begin{tabular}{cccccc}
 \PH & \PH & Exclusive & No. of & $m_{\PH}$ range & $m_{\PH}$ \\
 decay mode & production & final states & channels & [\GeVns{}] & resolution \\
\hline
$\PW\PW \to \ell\nu\ell\nu$      & 0/1-jets  &  ((\Pe\Pe, \Pgm\Pgm), \Pe\Pgm) + (0 or 1 jets) &  4         & 145--600         & 20\%  \\
$\PW\PW \to \ell\nu\ell\nu$      & VBF tag    &   ((\Pe\Pe, \Pgm\Pgm), \Pe\Pgm) + $(\mathrm{jj})_\mathrm{VBF}$  &  2    & 145--600         & 20\%  \\
$\PW\PW \to \ell\nu\cPq\cPq$          & untagged  &  ($\Pe\nu$, $\Pgm\nu$) + ((jj)$_{\PW}$ with 0 or 1 jets) & 4    & 180--600         & 5--15\%  \\
\hline
$\cPZ\cPZ \to 2\ell 2\ell'$             & inclusive  &  4\Pe, 4\Pgm, 2\Pe2\Pgm                                                                          & 3         & 145--1000         & 1--2\%   \\
      &   &  (\Pe\Pe, \Pgm\Pgm) + (\tauh\tauh, \taue\tauh, \taumu\tauh, \taue\taumu) & 8         & 200--1000         & 10--15\%  \\
$\cPZ\cPZ \to 2\ell 2\cPq$          & inclusive  &  (\Pe\Pe, \Pgm\Pgm) + ($(\mathrm{jj})_\cPZ$ with 0, 1, 2 b-tags)                                           & 6         & 200--600 & 3\% \\
$\cPZ\cPZ \to 2\ell 2\nu$        & untagged  &  (\Pe\Pe, \Pgm\Pgm) + 0, 1, 2 non-VBF jets                                        & 6       & 200--1000         & 7\%   \\
$\cPZ\cPZ \to 2\ell 2\nu$        & VBF tag    &  (\Pe\Pe, \Pgm\Pgm) + $(\mathrm{jj})_\mathrm{VBF}$                                                           & 2         & 200--1000         & 7\%   \\
\hline
\end{tabular}
\end{center}
\end{table*}

The results of all analyses are finally combined following the prescription developed by the ATLAS and CMS Collaborations
in the context of the LHC Higgs Combination Group~\cite{LHC-HCG-Report}, as described in
Ref.~\cite{Chatrchyan:2012tx}, taking into account the systematic uncertainties and their correlations.

\subsection{\texorpdfstring{$\PH \to \PW\PW \to \ell\nu\ell\nu$}{H to WW to ell nu ell nu}}
\label{sec:wwlvlv}

In this channel, the Higgs boson decays to two $\PW$ bosons, both of which decay leptonically, resulting in a signature with two isolated, oppositely charged, high-$\pt$ leptons (electrons or muons) and large $\MET$ due to the undetected neutrinos.
The analysis is very similar to that reported in Ref.~\cite{CMSobservation125,CMSlongpaper}, but additionally uses an improved Higgs boson mass lineshape model, and uses an MVA shape analysis~\cite{Chatrchyan:2012ty} for data taken at $\sqrt{s}=8\TeV$.
Candidate events must contain two reconstructed leptons with opposite charge, with $\pt>20\GeV$ for the leading lepton, and $\pt>10\GeV$ for the
second lepton. Only electrons (muons) with $\abs{\eta}<$2.5 (2.4) are considered in this channel.

Events are classified into three mutually exclusive categories, according to the number of reconstructed jets
with $\pt>30\GeV$ and $\abs{\eta}<$4.7. The categories are characterized by different signal yields and signal-to-background ratios. In the following these are referred to as 0-jet, 1-jet, and 2-jet samples. Events with more than two jets are considered only if they are consistent with the VBF hypothesis and therefore
must not have additional jets in the pseudorapidity region
between the highest-$\pt$ jets. Signal candidates are further divided into same-flavor leptons ($\Pep\Pem$, $\mumu$) and different-flavor leptons ($\Pe^\pm\Pgm^\mp$) categories. The bulk of the signal arises through direct $\PW$ decays to electrons or muons, with the
small contribution from $\PW \to \tau\nu \to \ell\mathrm{+X}$ decays implicitly included. The different-flavor lepton 0-jet and 1-jet categories are analysed with a multivariate technique, while all others make use of sequential selections.

{\tolerance=500 In addition to high-$\pt$ isolated leptons and minimal jet activity, $\MET$ is expected to be present in signal events, but
generally not in background.
For this channel, a $\METpr$ variable is employed.
The $\METpr$ is defined
as (i) the magnitude  of the $\vecEtm$
 component transverse to the closest lepton, if $\Delta \phi(\ell, \vecEtm ) < \pi/2$,
or (ii) the magnitude of the $\vecEtm$ otherwise.
This observable more efficiently rejects $\dytt$ background events in which
the $\vecEtm$
is preferentially aligned with the leptons,
and
$\dyll$ events with mismeasured
$\vecEtm$.
Since the $\METpr$ resolution is degraded as pileup increases,
the minimum of two different observables is used: the first includes all particle candidates in the
event, while the second uses only the charged
particle candidates associated with the primary vertex.
Events with $\METpr$ above 20\GeV are selected for this analysis.\par}

The backgrounds are suppressed using techniques described in Ref.~\cite{CMSobservation125,CMSlongpaper}. Top quark background is controlled with a top-quark-tagging technique based on soft muon and b-jet tagging~\cite{BTV}. A minimum dilepton transverse momentum ($\pt^{\ell\ell}$) of 45\GeV is required, in order to reduce the $\Wjets$ background. Rejection of events with
a third lepton passing the same requirements as the two selected leptons reduces both $\WZ$ and $\wgamma^*$ backgrounds.
The background from low-mass resonances is rejected by requiring a dilepton mass $\mll>12\GeV$.

The Drell--Yan process produces same-flavor lepton pairs ($\Pep\Pem$ and $\Pgmp\Pgmm$) and therefore
additional requirements are
applied for the same-flavor final state. Firstly, the resonant component of the Drell--Yan background
is rejected by requiring a dilepton mass outside a 30\GeV window centered on the $\cPZ$-boson mass. The remaining
off-peak contribution is further suppressed by requiring $\METpr>45\GeV$.
For events with two jets, the dominant source of misreconstructed \MET is the mismeasurement of
the hadronic recoil, and optimal performance is obtained by requiring $\MET>45\GeV$. Finally, the momenta of the dilepton system and of the most energetic jet must not be back-to-back in the transverse plane. These selections
reduce the Drell--Yan background by three orders of magnitude, while rejecting less than 50\% of the signal.

\begin{figure}[htbp]
\centering
   \includegraphics[width=0.49\textwidth]{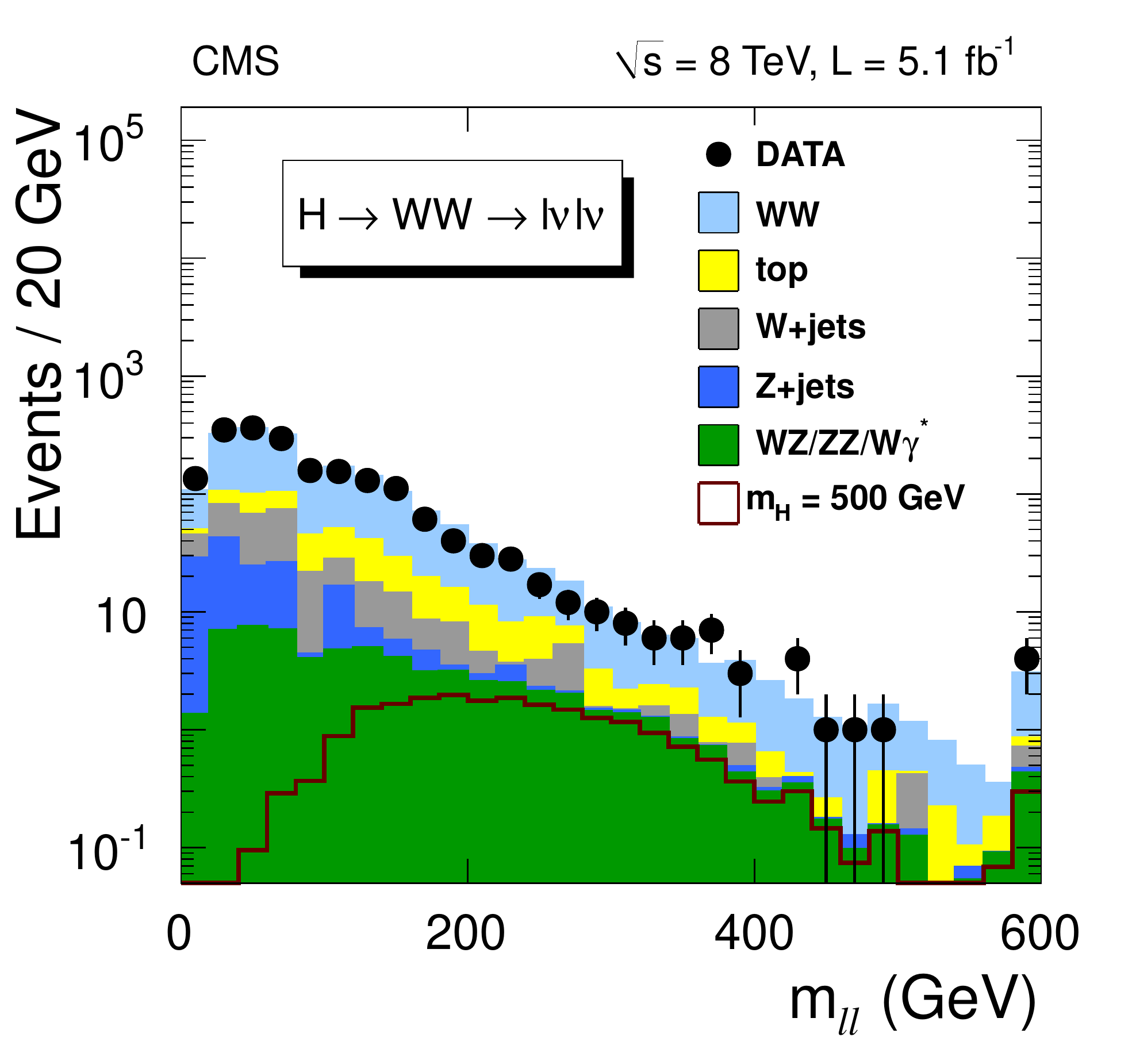}
   \includegraphics[width=0.49\textwidth]{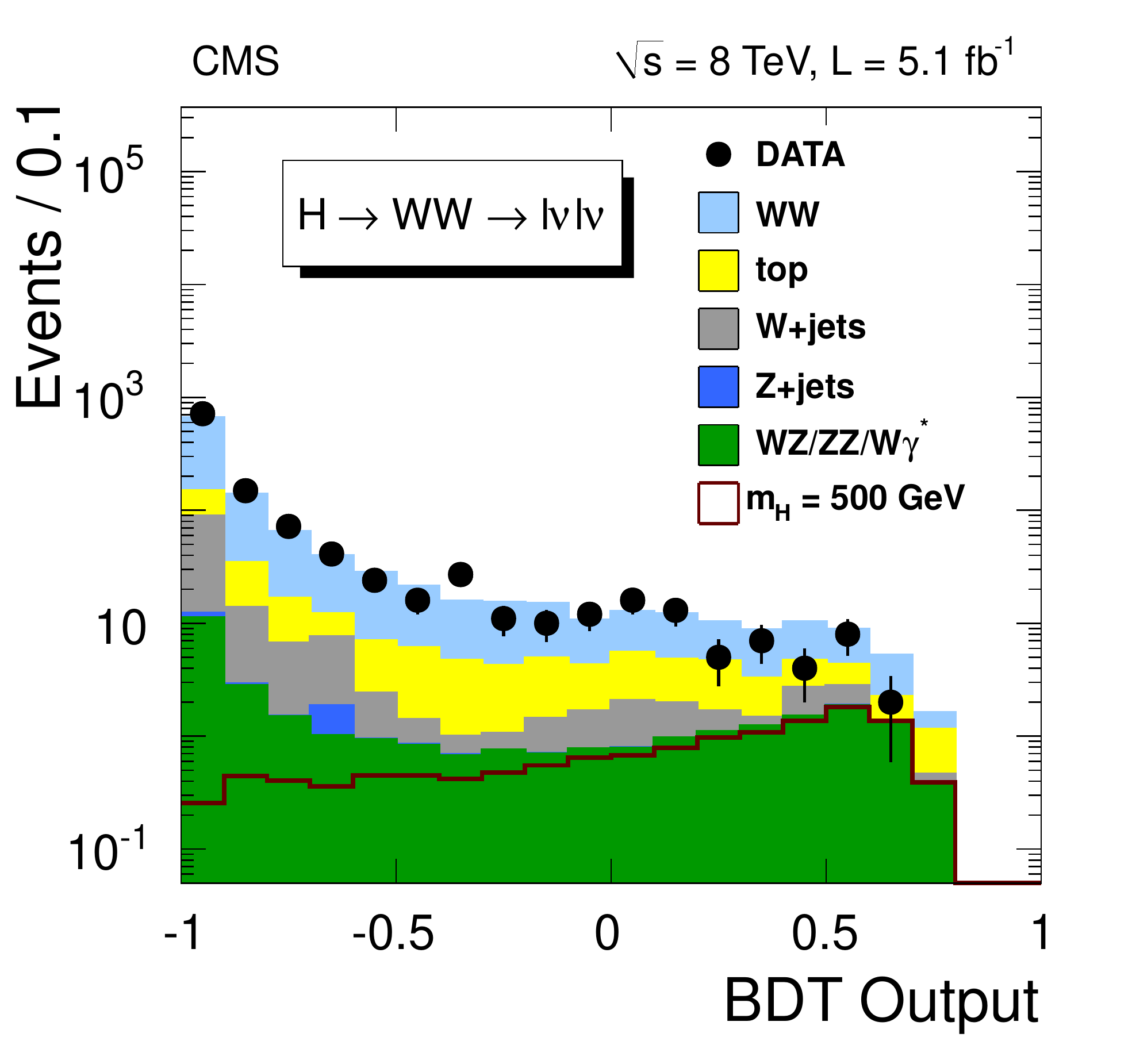}
   \caption{(\cmsLeft) Distributions of $m_{\ell\ell}$
   in the 0-jet different-flavor category of the $\PW\PW \to \ell\nu\ell\nu$ channel for data (points with
   error bars), for the main backgrounds (stacked histograms), and for
   a SM Higgs boson signal with $\mH= 500\GeV$.
   The standard preselection is applied.  (\cmsRight) BDT-classifier
   distributions for signal and background events for a SM
   Higgs boson with $\mH=500\GeV$ and for the main backgrounds 
   in the 0-jet different-flavor category after
   requiring $80 < \mt^{\ell\ell,\MET} < 500\GeV$ and $m_{\ell\ell} < 500\GeV$.}
\label{fig:hww2l2n_bdt_500}
\end{figure}

These requirements form the set of ``preselection'' criteria. The preselected sample is dominated by non-resonant $\WW$ events. Figure~\ref{fig:hww2l2n_bdt_500}(\cmsLeft) shows an example of
the $m_{\ell\ell}$ distribution for the 0-jet different-flavor-leptons category after the preselection. The data are well reproduced by the simulation.
To enhance the signal-to-background ratio, loose $\mH$-dependent requirements are applied on $\mll$ and the transverse mass, given by:
\begin{equation*}
\mt^{\ell\ell,\MET} = \sqrt{2 \pt^{\ell\ell} \MET (1-\cos\delphimetll)},
\end{equation*}
where $\delphimetll$ is the difference in azimuth between $\vecEtm$ and $\vecPtell$.
After preselection, a multivariate technique is employed for the different-flavor final state in the 0-jet
and 1-jet categories. In this approach, a boosted decision tree (BDT)~\cite{tmva} is trained for each Higgs boson mass hypothesis and jet category to discriminate signal from background.

The multivariate technique employs the variables used in the preselection and additional observables including
$\Delta R_{\Lep\Lep}$ between the leptons and the $\mt^{\ell\ell,\MET}$. For the 1-jet category the $\delphimetll$ and
azimuthal angle between the $\vecPtell$ and the jet are also used. The BDT classifier distributions for $\mH=500\GeV$ are shown in Fig.~\ref{fig:hww2l2n_bdt_500}(\cmsRight) for the 0-jet different-flavor category. BDT training is performed using $\Hww$ as signal and non-resonant $\WW$ as background. 
The sum of templates for the signal and background 
are fitted to the binned observed BDT distributions.

The 2-jet category is optimized for the VBF production mode \cite{Ciccolini:2007jr,Ciccolini:2007ec,Arnold:2008rz,Cahn:1987}, for which the cross section is roughly ten times smaller than for the gluon fusion mode. Sequential selections are employed for this category.
The main requirements for selecting the VBF-type events are on the mass of the dijet system, $\mjj > 450\GeV$, and on the angular separation of the two jets $|\Delta \eta_{\mathrm{jj}}| > 3.5$. An $\mH$-dependent requirement on the dilepton mass
is imposed, as well as other selection requirements that are independent of the Higgs boson mass hypothesis.

The normalization of the background contributions relies on data whenever possible and exploits a combination
of techniques~\cite{CMSobservation125,CMSlongpaper}. The $\ttbar$ background is estimated by extrapolation from the observed
number of events with the b-tagging requirement inverted. The Drell--Yan background measurement is based on extrapolation
from the observed number of $\Pep\Pem$, $\Pgmp \Pgmm$ events with the $\cPZ$-veto requirement inverted. The background
of $\PW+\text{jets}$ and QCD multi-jet events is estimated by measuring the number of events with one lepton passing a loose
requirement on isolation. The probability for such loosely-isolated non-genuine leptons to pass the tight isolation criteria is measured in
data using multi-jet events. The non-resonant $\WW$ contribution is estimated from simulation.

Experimental effects, theoretical predictions, and the choice of event generators are considered as sources of
systematic uncertainty, and their impact on the signal efficiency is assessed. The impact on the kinematic distributions is also considered for the BDT analysis. The overall signal yield uncertainty is estimated to be about 20\%, and is
dominated by the theoretical uncertainty associated with missing higher-order QCD corrections and PDF uncertainties, 
estimated following the PDF4LHC recommendations~\cite{Alekhin:2011sk,Botje:2011sn,Lai:2010vv,Martin:2009iq,Ball:2011mu}. 
The total uncertainty on the background estimation in the $\Hww$ signal region is about 15\% and is dominated by the
statistical uncertainty on the observed number of events in the background control regions.

\begin{figure}[htbp]
  \centering
  \includegraphics[width=\cmsFigWidth]{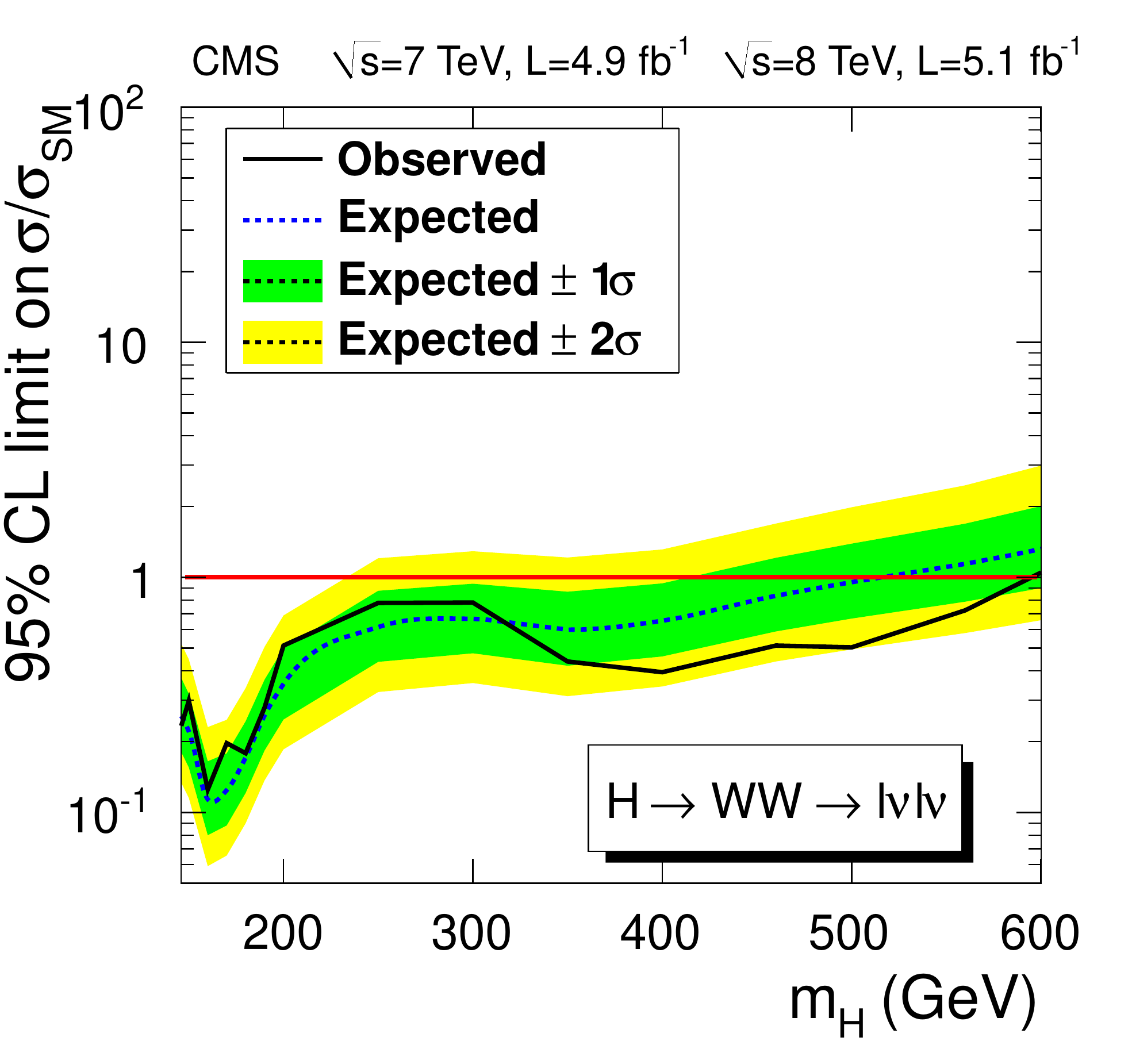}
  \caption{\label{fig:hwwlvlvlim}Observed (solid line) and expected
    (dashed line) 95\% CL upper limit on the ratio of the product of production cross
    section and branching ratio to the SM expectation for the Higgs boson obtained using
    the asymptotic CL$_{\textrm{S}}$ technique~\cite{Junk:1999kv,Read1} in the $\PW\PW \to \ell\nu\ell\nu$ channel.
    The 68\% ($1\sigma$) and 95\% ($2\sigma$) CL
    ranges of expectation for the background-only model are also shown
    with green and yellow bands, respectively. The horizontal solid line at unity
    indicates the SM expectation. Color figure online.
}
\end{figure}

After applying the final selections, no evidence of a SM-like Higgs boson is observed over the mass range considered in this paper. Upper limits are derived on the ratio of the product of the Higgs boson production cross section and the $\Hi \to \WW$ branching fraction, $\sigma_{\Hi} \times \mathcal{B}(\Hi \to \WW)$, to the SM expectation. The observed and expected upper limits at 95\% confidence
level (CL) with all categories combined are shown in Fig.~\ref{fig:hwwlvlvlim}.
The contribution of the 2-jet category to the expected limits is approximately 10\%.

\subsection{\texorpdfstring{$\PH \to \PW\PW \to \ell\nu \cPq\cPq$}{H to WW to ell nu qq}}

The $\WW$ semileptonic channel has the largest branching fraction of all the channels presented in this paper.
Its advantage over the fully leptonic final state is that it has a reconstructable Higgs boson mass peak~\cite{intro2}. This
comes at the price of a large $\PW+\text{jets}$ background. The level to which this background can be controlled
largely determines the sensitivity of the analysis. This is the first time
CMS is presenting
 a measurement in this decay channel.

The reconstructed electrons (muons) are required to have
$\PT>35\,(25)\GeV$, and are restricted to $\abs{\eta}<2.5\, (2.1)$.
The jets are required to have
$\PT>30\GeV$ and $\abs{\eta}<2.4$, and not to
overlap with the leptons,
with the overlap determined by a cone around the lepton axis of radius $\Delta R = 0.3$.
Events with electrons and muons, and with exactly two or three jets are analysed separately,
giving four categories in total.
The two highest-$\PT$ jets are assumed to arise from the hadronic decay of the $\PW$ candidate. According to simulation, in the case of 2\,(3) jet events, the correct jet-combination rate varies from 68 (26)\% for $\mH = 200\GeV$ to 88\,(84)\% for
$\mH = 600\GeV$. For low $\mH$ values jets produced in initial or final state radiation are often
more energetic than jets from $\PW$ decay, therefore in 3 jet events the correct jet-combination
rate decreases quickly with decreasing $\mH$.
Events with an incorrect dijet combination result in a broad non-peaking background in the $m_{\WW}$ spectrum.

{\tolerance=500 The leptonic $\PW$ candidate is reconstructed from the $(\ell, \MET)$ system. Events are required to have $\MET > 30\,(25)\GeV$ for the electron (muon) categories. To reduce the background from processes that do not
contain $\PW\to\ell\nu$ decays, requirements of $m_{\mathrm{T}}^{\ell,\MET}>30\GeV$ and
\breakhere$\abs{\Delta\phi_{\textrm{leading jet,\MET}}} > 0.8\,(0.4)$ are imposed for electrons (muons). 
The $m_{\mathrm{T}}^{\ell,\MET}$ is defined as
\begin{equation*}
\mt^{\ell,\MET} = \sqrt{2 \pt^{\ell} \MET (1-\cos\delphimetl)},
\end{equation*}
where $\delphimetl$ is the difference in azimuth between $\vecEtm$ and $\vecPtel$.
These criteria reduce the QCD multijet background, for which in many cases the $\MET$ is generated by a mismeasurement of a jet energy.\par}

To improve the $m_{\WW}$ resolution, both $\PW$ candidates are constrained in a kinematic fit to the $\PW$-boson
mass to within its known width. For the $\PW \to \cPq\cPq$ candidate the fit 
uses the 
four-momenta of the
two highest-$\pt$ jets. For the $\PW \to \ell\nu$ candidate the $\MET$ defines the transverse energy of the neutrino and the  longitudinal component of the neutrino momentum, $p_z$, is unknown. 
The ambiguity is resolved by taking the solution that yields the smaller $\abs{p_z}$ value for the neutrino.
According to simulation over 85\% of signal events receive a correct $\abs{p_z}$ value, thus
improving the mass resolution, especially at low $\mH$. 

To exploit the differences in kinematics between signal and background events, a likelihood discriminant is constructed
that incorporates a set of variables that best distinguishes the Higgs boson signal from the $\PW+\text{jets}$ background.
These variables comprise five angles between the Higgs boson decay products, that describe the Higgs boson production kinematics~\cite{Gao:2010qx}; the $\PT$ and rapidity of the $\WW$ system; and the lepton charge.
The likelihood discriminant is optimized with dedicated simulation samples for several discrete Higgs boson mass hypotheses,
for each lepton flavor ($\Pe$, $\Pgm$) and for each jet multiplicity (2-jet, 3-jet) independently. Four different optimizations are therefore obtained per mass hypothesis. For each of them, events are retained if they
survive a simple selection on the likelihood discriminant, chosen in order to optimize the expected limit for the Higgs boson production  cross section.

To simultaneously extract the relative normalizations of all background components in the signal region, an
unbinned maximum likelihood fit is performed on the invariant mass distribution of the dijet system, $\mjj$.
The fit is performed independently for each Higgs boson mass hypothesis. The signal region corresponding to the $\PW$
mass window, $65 < \mjj < 95\GeV$, is excluded from the fit.
The mass window corresponds to approximately twice the dijet mass resolution.
The shape of the $\mjj$ distribution for the
$\PW+\text{jets}$ background is determined by simulation.
The overall normalization of the \PW+jets component is allowed to vary in the fit. The shapes for
other backgrounds (electroweak diboson, \ttbar, single top quark, and Drell--Yan plus jets) are based on simulation, and their
normalizations are constrained to theoretical predictions, within the corresponding uncertainties.
The multijet background normalization is estimated from data by relaxing lepton isolation and identification requirements.
Its contribution to the total number of events is evaluated from a separate two-component likelihood fit to the $m_{\mathrm{T}}^{\ell,\MET}$ distribution, and constrained in the $\mjj$ fit according to this fraction within
uncertainties. For electrons, the multijet fraction accounts for several percent of the event sample, depending on the
number of jets in the event, while for muons it is negligible.

\begin{figure}[htbp]
  \centering
  \includegraphics[width=0.48\textwidth]{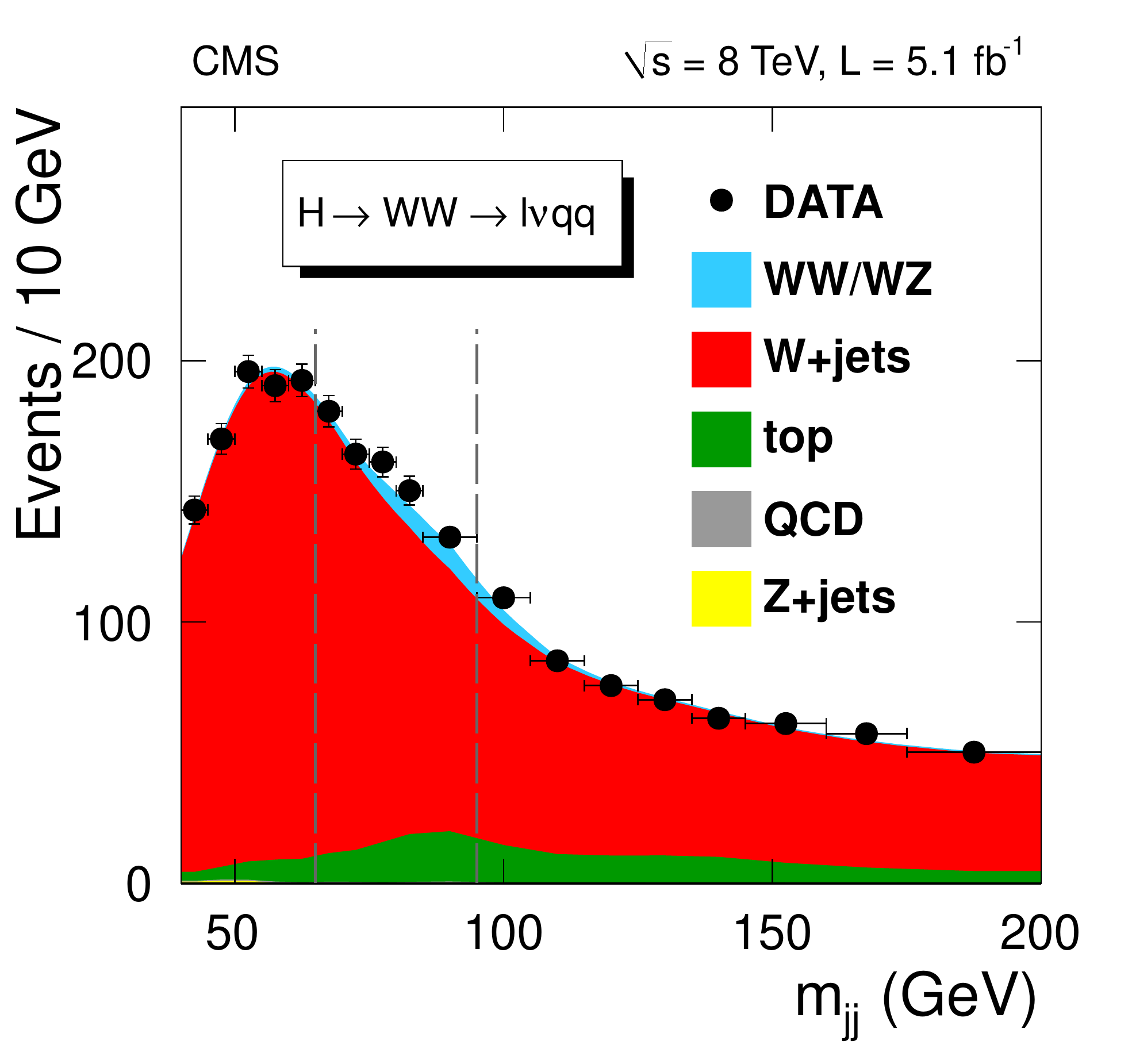}
  \includegraphics[width=0.48\textwidth]{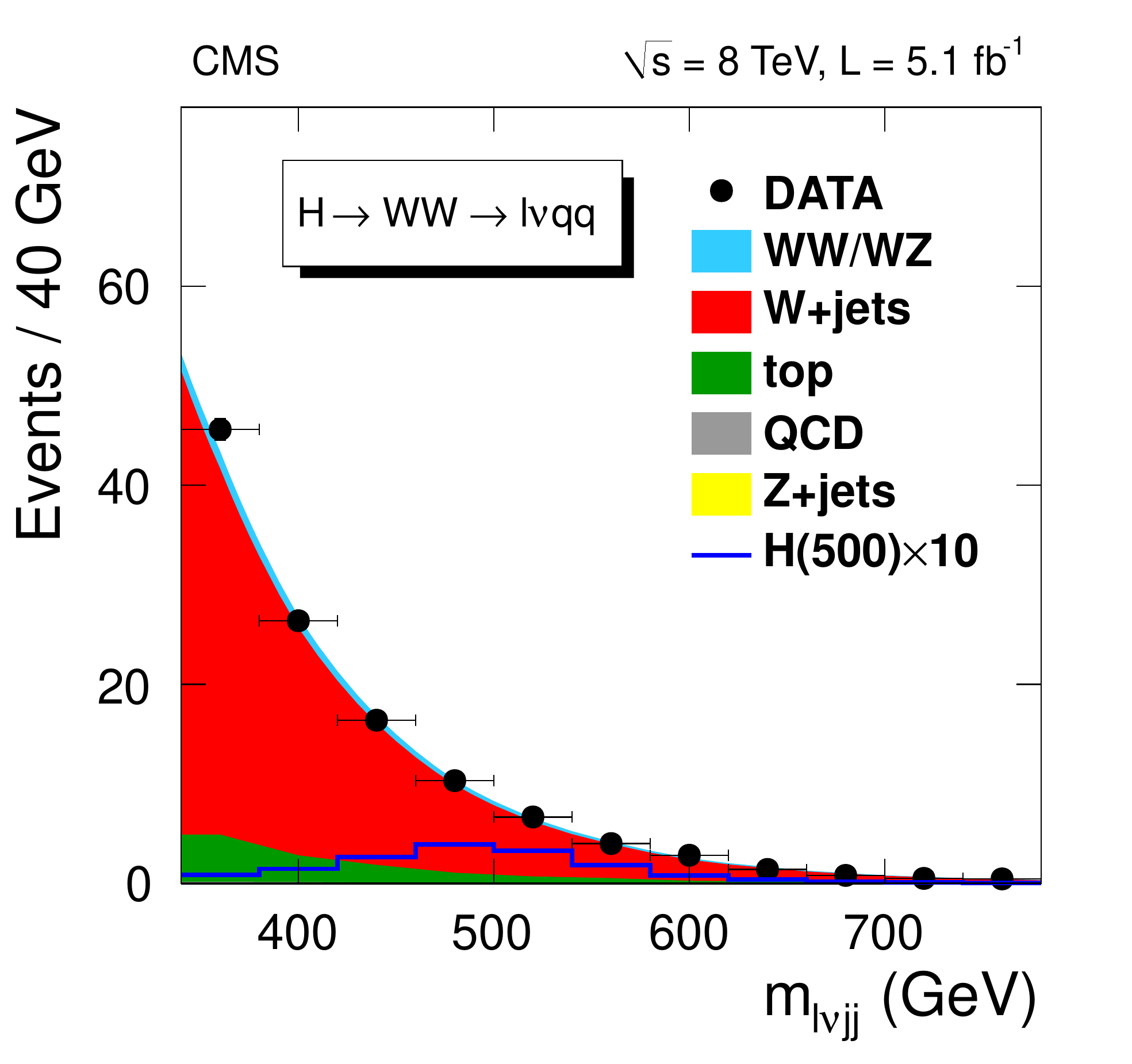}
  \caption{\label{fig:lvjjfits} Invariant mass distributions for the
    $\mH=500\GeV$ mass hypothesis, (\Pgm, 2 jets) category in the
    $\PH \to \PW\PW \to \ell\nu \cPq\cPq$ channel.
    (\cmsLeft) The dijet invariant mass
    distribution with the major background contributions.
    The vertical lines correspond to the signal region of this analysis
    $65 < \mjj
    < 95\GeV$.
    (\cmsRight) The $\WW$ invariant mass distribution with the major background
    contributions in the signal region.
    }
\end{figure}

Limits are established based on the measured invariant mass of the $\WW$ system, $m_{\ell\nu\mathrm{jj}}$. 
The $m_{\ell\nu\mathrm{jj}}$ shape for the major background, \PW+jets,
is extracted from data as a linear combination of the shapes measured in two signal-free sideband
regions of $\mjj$ ($55 <\mjj < 65\GeV$,
$95 < \mjj < 115\GeV$). The relative fraction of the two sidebands is determined
through simulation, separately for
each Higgs boson mass hypothesis, by minimizing the $\chi^2$ between the interpolated $m_{\ell\nu\mathrm{jj}}$ shape in the
signal region and the expected one. The $m_{\ell\nu\mathrm{jj}}$ shape for multijet background events is
obtained from data with the procedure described above.
All other background categories use the $m_{\ell\nu\mathrm{jj}}$ shape
from simulation.
The $\mjj$ and $m_{\ell\nu\mathrm{jj}}$ distributions with final background estimates are
shown in Fig.~\ref{fig:lvjjfits}, with
selections optimized
for a $500\GeV$ Higgs boson mass hypothesis, for the (\Pgm, 2 jets) category.
The final background $m_{\ell\nu\mathrm{jj}}$ distribution is obtained by summing
up all the individual contributions and smoothing it with an exponential function.
The shapes of the $m_{\ell\nu\mathrm{jj}}$ distribution
 for total background, signal and data for each
mass hypothesis and
event category are binned, with bin size approximately equal to the mass resolution,
 and fed as input to the limit-setting procedure.

The largest source of systematic uncertainty on the background
is due to the uncertainty in the
shape of the
 $m_{\ell\nu\mathrm{jj}}$ distribution of the total background.
The shape uncertainty is derived by varying the parameters of the exponential
fit function  up and down by 
one standard deviation.
The only other uncertainty assigned to background is the normalization uncertainty from the $\mjj$ fit. Both of these
uncertainties are estimated from data.
The dominant
systematic uncertainties on the signal include theoretical uncertainties for the cross section (14--19\% for gluon fusion)~\cite{LHCHiggsCrossSectionWorkingGroup:2011ti} and on jet energy
 scale (4--28\%), as well as the
efficiency of the likelihood selection (10\%). The latter effect is computed by taking the relative difference in efficiency
between data and simulation using a control sample of top-quark pair events in data. These events are good proxies for the signal,
since in both cases the primary production mechanism is gluon fusion, and the semi-leptonic final states contain
decays of two W bosons.

\begin{figure}[htbp]
  \centering
  \includegraphics[width=\cmsFigWidth]{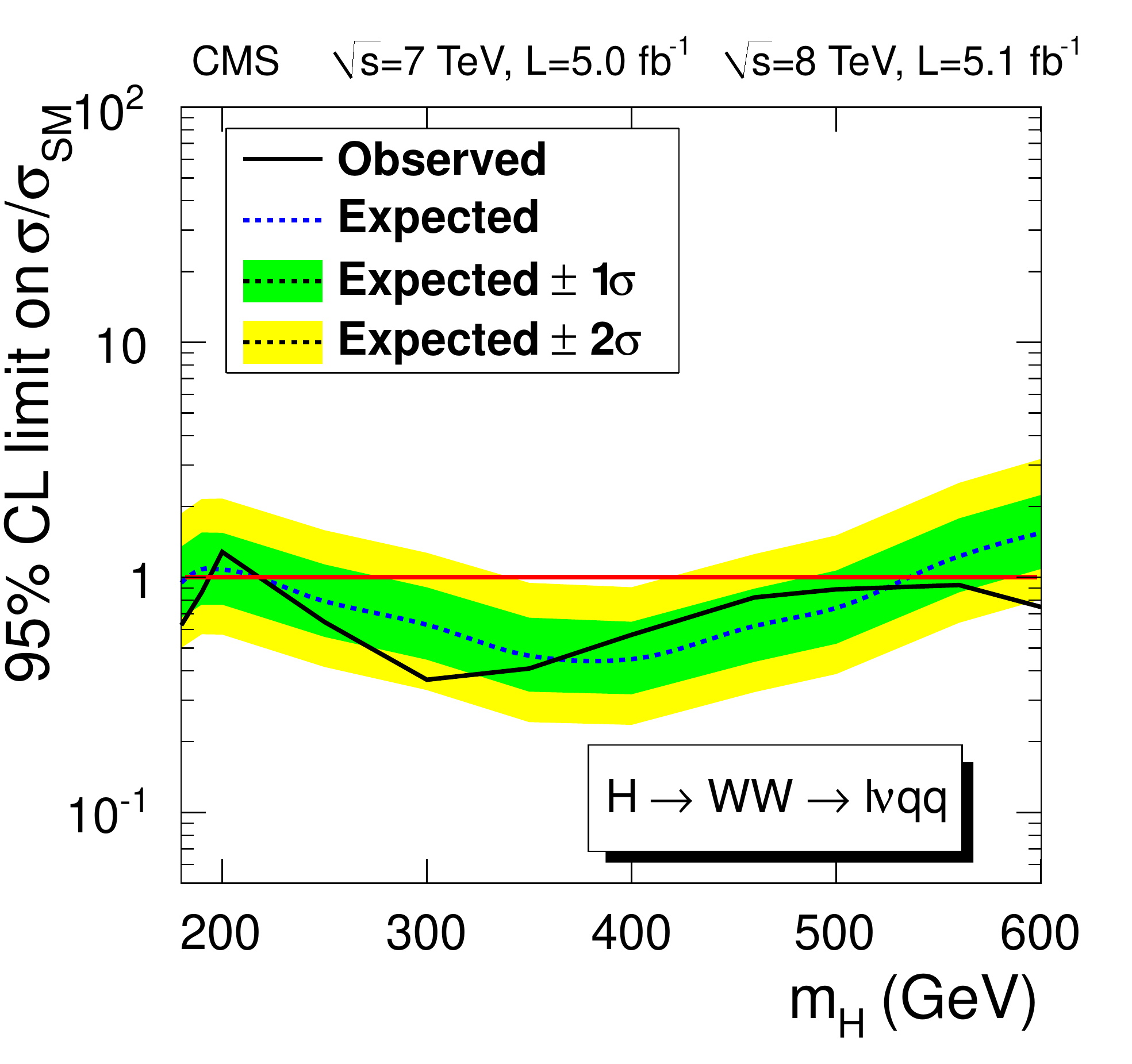}
  \caption{\label{fig:hwwlvjjlim} Observed (solid line) and expected
  (dashed line) 95\% CL upper limit on the ratio of the product of production cross
  section and branching fraction to the SM expectation for the Higgs boson in the WW
  semileptonic channel.}
\end{figure}

The upper limits on the ratio of the production cross section for the Higgs boson compared to the SM expectation are
presented in Fig.~\ref{fig:hwwlvjjlim}.

\subsection{\texorpdfstring{$\PH \to \ZZ \to 2\ell 2\ell '$}{H to ZZ to 2 ell 2 ell'}}

This analysis seeks to identify Higgs boson decays to a pair of $\cPZ$ bosons, with both decaying to a pair of
leptons. This channel has extremely low background, and the presence of four leptons in the
final state allows reconstruction and isolation requirements to be loose. Due to very good mass resolution and  high efficiency of the
selection requirements, this channel is one of the major discovery channels at both low and high Higgs boson masses.
A detailed description of this analysis may be found in~\cite{CMSobservation125,CMSlongpaper,Chatrchyan:2012dg,Chatrchyan:2012hr}.

Events included in the analysis contain $\cPZ$ candidates formed from a pair of leptons of the same flavor and
opposite charge. Electrons (muons, $\tauh$) are required to be isolated, to originate from the primary vertex, and
to have $\pt > 7\, (5,\, 20)\GeV$ and $\abs{\eta}<2.5\, (2.1,\, 2.3)$.
The event selection procedure results in mutually exclusive sets of Z candidates
in the $\PH\to 2\ell 2\ell$ and $\PH\to 2\ell2\Pgt$ channels, with the former identified first.

For the $2\ell2\ell$ final state, the lepton pair with invariant mass closest to the nominal $\cPZ$ boson  mass, denoted
$\cPZ_1$, is identified and retained if it satisfies $40 < m_{\cPZ_1} < 120\GeV$. The second $\cPZ$ candidate  is then
constructed from the remaining leptons in the event, and is required to satisfy $12 < m_{\cPZ_2} < 120\GeV$. If
more than one $\cPZ_2$ candidate remains, the ambiguity is resolved by choosing the leptons of highest $\pt$. Amongst
the four candidate decay leptons, it is required that at least one should have $\PT > 20\GeV$, and that another
should have $\PT > 10\GeV$. This requirement ensures that selected events correspond to the high-efficiency
plateau of the trigger.

For the $2\ell2\Pgt$ final state, events are required to have one $\cPZ_1 \to \ell^+\ell^-$ candidate, with one
lepton having $\PT > 20\GeV$ and the other $\PT > 10\GeV$, and a $\cPZ_2 \to \Pgt^+\Pgt^-$, with $\Pgt$ decaying
to $\Pgm, \Pe$ or hadrons. The leptons from $\Pgt$ leptonic decays are required to have $\PT > 10\GeV$. The invariant
mass of the reconstructed $\cPZ_1$ is required to satisfy $60 < m_{\ell\ell} < 120\GeV$, and that of the $\cPZ_2$ to
satisfy $ m_{\Pgt\Pgt} < 90\GeV$, where $m_{\Pgt\Pgt}$ is the invariant mass of the visible $\tau$-decay products.

Simulation is used to evaluate the expected non-resonant $\cPZ\cPZ$ background as a
function of $m_{2\ell2\ell'}$.
The cross section for \ZZ\ production at NLO is
calculated with \textsc{mcfm}~\cite{MCFM,Campbell:1999ah,Campbell:2011bn}. The theoretical uncertainty on the
cross-section is evaluated as a function of $m_{2\ell2\ell'}$, by varying the QCD renormalization and factorization
scales and the PDF set, following the PDF4LHC recommendations.
The uncertainties associated
with the QCD and PDF scales for each final state are on average 8\%. The number of predicted $\ZZ \to 2\ell 2\ell'$
events and their associated uncertainties, after the signal selection, are given in Table~\ref{tab:SelectYields}.

To allow estimation of the $\ttbar$, $\cPZ+\text{jets}$, and $\PW\cPZ+\text{jets}$ reducible backgrounds
a $\cPZ_1+\ell_{\mathrm{ng}}$
control region
is
defined,
with at least one loosely defined non-genuine lepton candidate, $\ell_\mathrm{ng}$,
in addition to a $\cPZ$ candidate.
To avoid possible contamination from $\PW\cPZ$ events, $\MET < 25\GeV$ is required.
This control region is used to determine the misidentification probability for $\ell_{\mathrm{ng}}$ to pass the final
lepton selections as a function of $\pt$ and $\eta$. 
To estimate the number of expected background events in the signal region,
$\cPZ_1+\ell^{\pm}\ell^{\mp}$, this misidentification probability is applied to two control regions,
$\cPZ_1+\ell^{\pm}\ell_{\text{ng}}^{\mp}$ and $\cPZ_1+\ell_{\mathrm{ng}}^{\pm}\ell_{\mathrm{ng}}^{\mp}$.
The contamination from $\PW\cPZ$ events containing a
genuine additional lepton is suppressed by requiring the imbalance of the measured energy deposition in the transverse
plane to be below 25\GeV.
The estimated reducible background yield in
the signal region is denoted as $\cPZ+$X in Table~\ref{tab:SelectYields}. The systematic uncertainties associated with the
reducible background estimate vary from 30\% to 70\%, and are presented in the table combined in quadrature with
the statistical uncertainties.

\begin{table*}[htbp]
\begin{center}
\topcaption{
Observed and expected background and signal yields for each final state in the
$\PH \to \ZZ \to 2\ell 2\ell '$ channel. For the $\cPZ$+X background, the estimations are based on data. 
The uncertainties represent the statistical and systematic uncertainties combined in quadrature.
}
\label{tab:SelectYields}
\begin{tabular}{lcccc}

Channel & $4\Pe$ & $4\Pgm$ & $2\Pe2\Pgm$ & $2\ell2\Pgt$  \\
\hline
\cPZ\cPZ\ background &  28.6  $\pm$  3.3 &  44.6  $\pm$  4.6  &  70.8  $\pm$  7.5 &  12.1 $\pm$ 1.5 \\ %
\cPZ+X                            &   $2.3 ^{ +  2.1 }_{ -  1.5 }$ &  $1.1 ^{ +  0.8 }_{ -  0.7 }$ &  $3.6 ^{ +  2.9 }_{ -  2.2 }$  & 8.9  $\pm$ 2.5  \\
\hline
All backgrounds    &   $30.9 ^{ +  3.9 }_{ -  3.6 }$ &  $45.7 ^{ +  4.7 }_{ -  4.7 }$ &  $74.4 ^{ +  8.0 }_{ -  7.8 }$ & 21.0 $\pm$ 2.9  \\
\hline
Observed  & 26 & 42  & 88 & 20 \\
\hline
$\mH = 350\GeV$ & 5.4  $\pm$  1.4  &  7.6  $\pm$  1.6  &  13.2  $\pm$  3.0 &  3.1 $\pm$ 0.8 \\
$\mH = 500\GeV$ & 1.9  $\pm$  0.9  &  2.7  $\pm$  1.2  &  4.6  $\pm$  2.1 &  1.4 $\pm$ 0.7 \\
\hline
\end{tabular}
\end{center}
\end{table*}

\begin{figure}[htbp]
\begin{center}
{\includegraphics[width=0.48\textwidth]{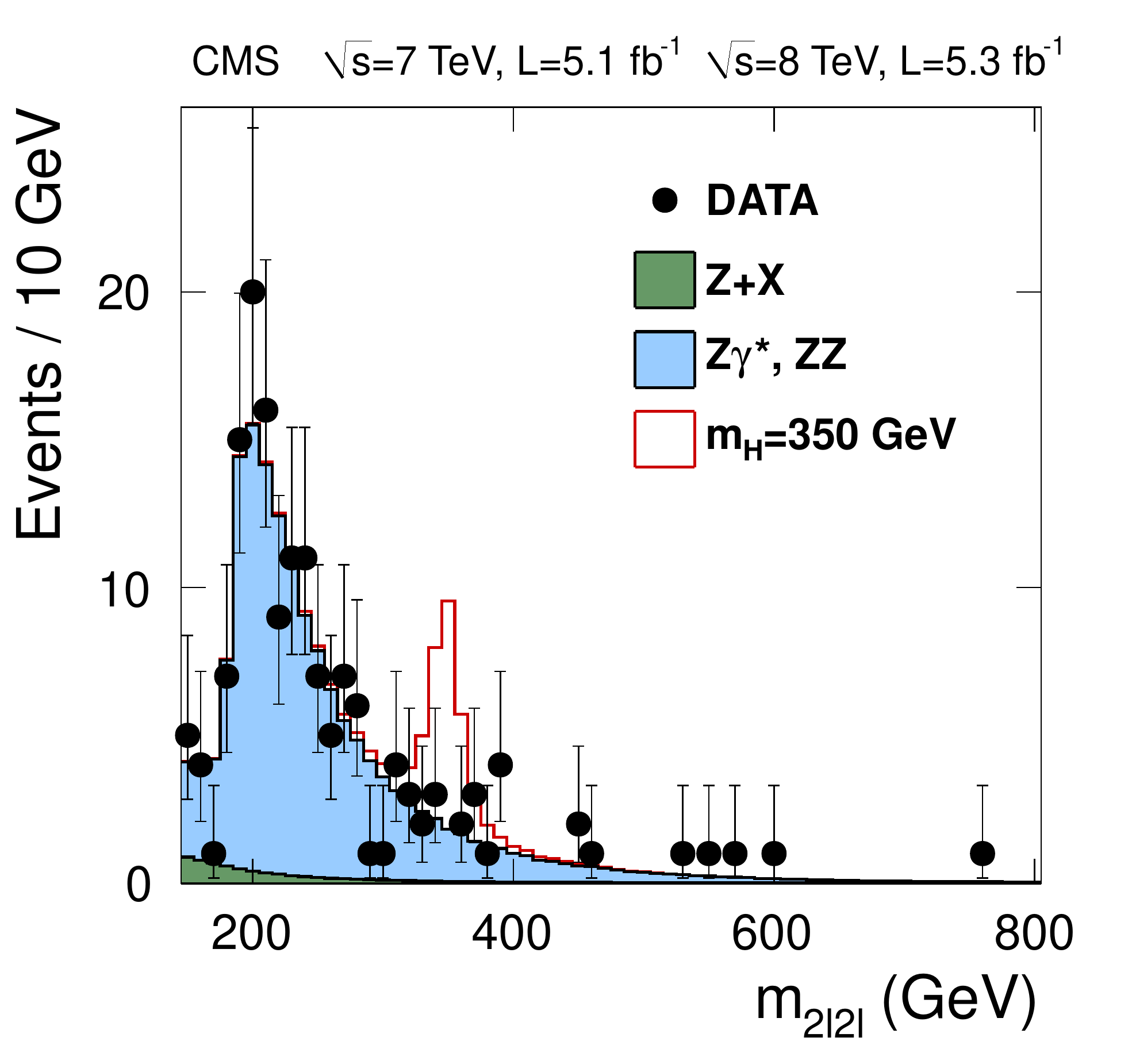}}
{\includegraphics[width=0.48\textwidth]{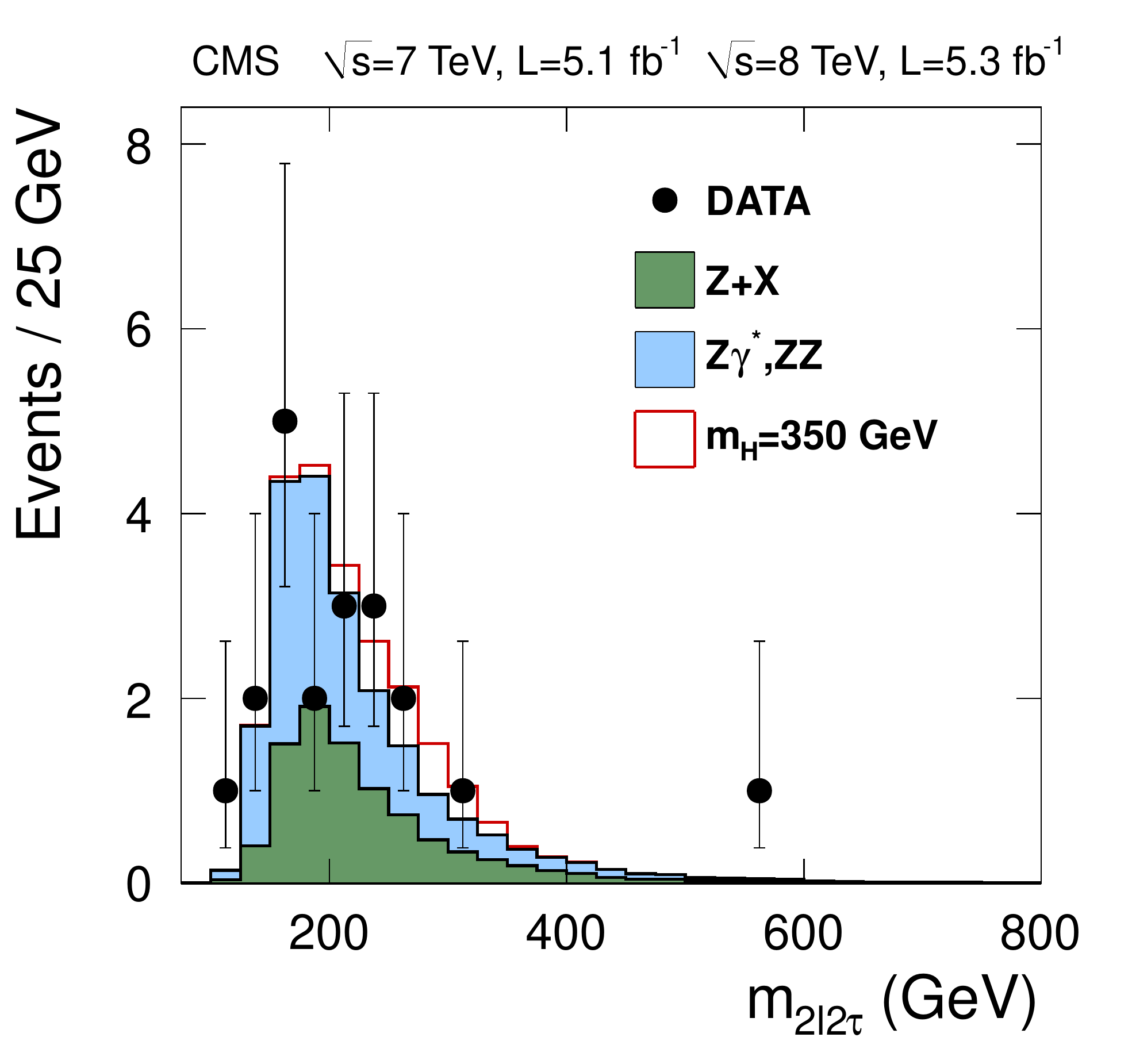}}
\caption{
Distribution of the four-lepton reconstructed mass for (\cmsLeft) the sum of
the $4\Pe$, $4\Pgm$, and $2\Pe2\Pgm$ channels, and for (\cmsRight) the sum over all
$2\ell2\tau$ channels.
Points represent the data, shaded histograms represent the background,
and unshaded histogram the signal expectations.
The reconstructed masses in $2\ell2\tau$ states
are shifted downwards with respect to the true masses by about
30\% due to the
undetected neutrinos in $\tau$ decays.
}
\label{fig:Mass4l-2l2tau}
\end{center}
\end{figure}

The reconstructed invariant mass distributions for $2\ell 2\ell'$
are shown in Fig.~\ref{fig:Mass4l-2l2tau} for
the combination of the $4\Pe$, $4\Pgm$, and $2\Pe2\Pgm$
final states in the \cmsLeft plot and for the combination of the $2\ell2\tau$ states
in the \cmsRight one.
The data are compared with the expectation from SM background processes.
The observed mass distributions are consistent with the SM background expectation.

\begin{figure}[htbp]
\begin{center}
\includegraphics[width=0.48\textwidth]{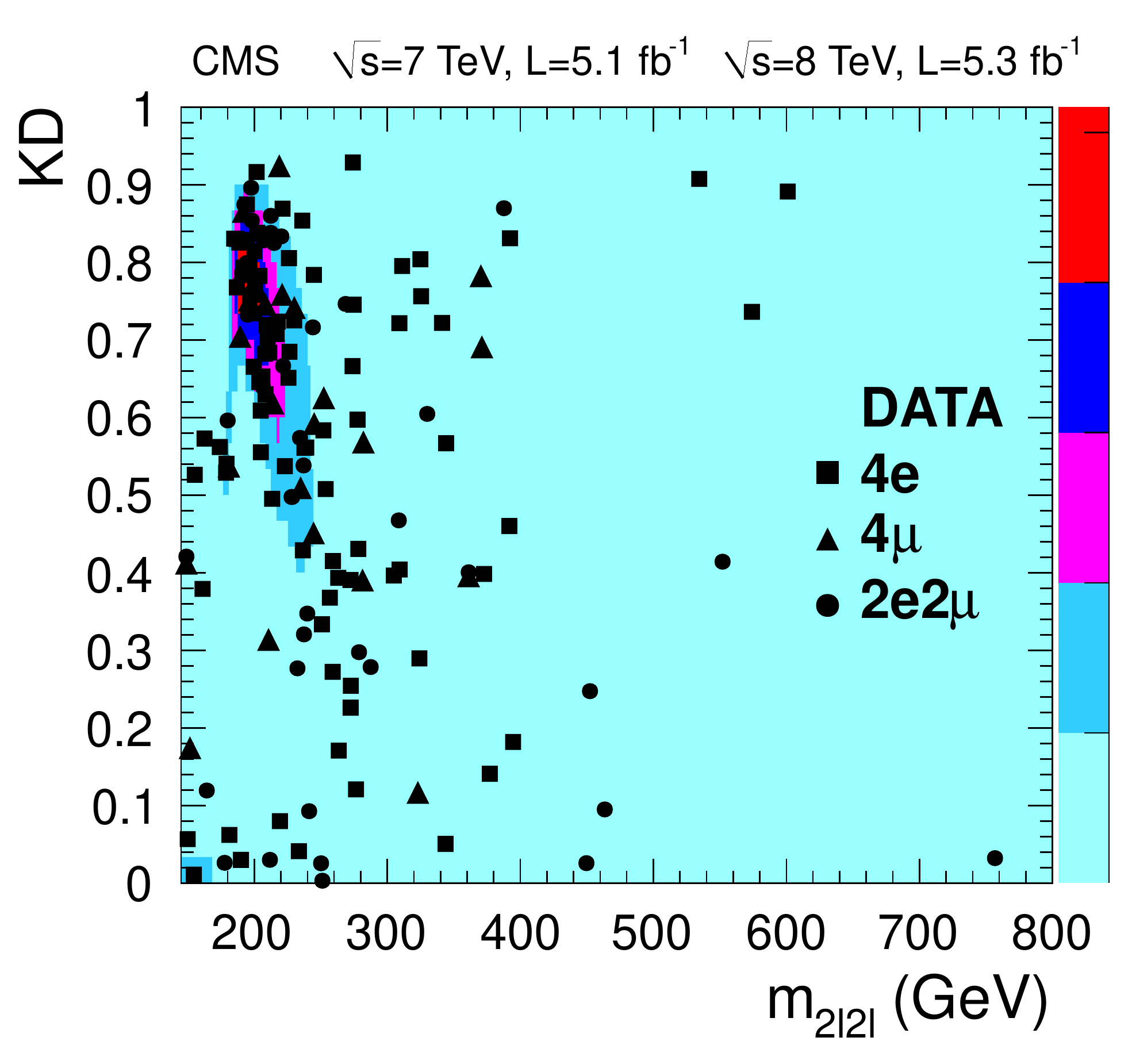}
\includegraphics[width=0.48\textwidth]{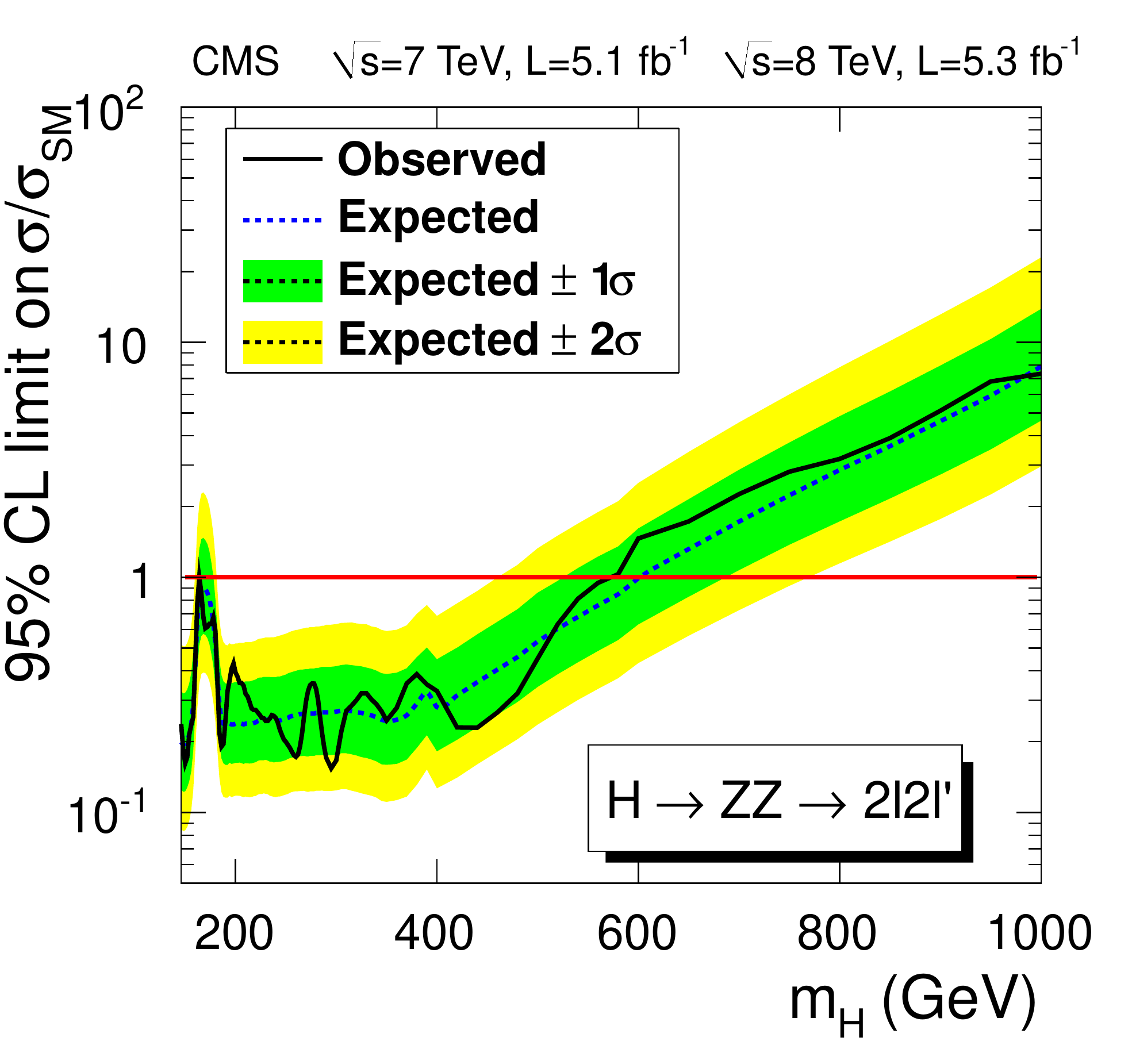}
\caption{ (\cmsLeft)
The distribution of events
selected in the $2\ell2\ell$
subchannels for the kinematic discriminant, KD, versus $m_{2\ell2\ell}$.
Events in the three final states are marked by filled symbols (defined in the legend).
The colored contours (with the measure on the color scale of the right axis) represent the expected relative density of background events.
(\cmsRight) Observed (solid line) and expected (dashed line) 95\% CL upper limits on the ratio of the product of the production cross section and
 branching fraction to the
SM expectation in the $\PH \to \ZZ \to 2\ell 2\ell'$ channel.
The 68\% ($1\sigma$) and 95\% ($2\sigma$) ranges of expectation  for the background-only model
are also shown with green and yellow bands, respectively. Color figure online.
}
\label{fig:KDvsM4lFullMass}
\end{center}
\end{figure}

The kinematics of the $\PH\to\ZZ \to 2\ell2\ell$ process, for a given invariant mass of the four-lepton system,
are fully described at LO by five angles and the invariant masses of the two lepton pairs~\cite{Cabibbo:1965zz,Gao:2010qx,DeRujula:2010ys}. A kinematic discriminant (KD), based on these seven variables, is constructed based on the probability ratio of the signal and background hypotheses~\cite{2l2qpaper}. The distribution of $\KD$ versus $m_{2\ell 2\ell}$ is shown in Fig.~\ref{fig:KDvsM4lFullMass}(\cmsLeft) for the selected event sample, and is consistent with the SM background expectation. The
two-dimensional KD-$m_{2\ell2\ell}$ distribution is used to set upper limits on the cross-section in the $2\ell2\ell$
channel. For the $2\ell2\tau$ final state, limits are set using the $m_{2\ell2\tau}$ distribution. The combined upper
limits from all channels are shown in Fig.~\ref{fig:KDvsM4lFullMass} (\cmsRight).

\subsection{\texorpdfstring{$\PH \to \ZZ \to 2\ell 2\cPq$}{H to ZZ to 2 ell 2q}}

This channel has the largest branching fraction of all $\PH \to \ZZ$ channels considered in this paper, but also a large background
contribution from $\cPZ+\text{jets}$ production. The hadronically-decaying $\cPZ$ bosons produce quark jets, with a large
fraction of heavy quarks compared to the background that is dominated by gluon and light quark jets.
This feature allows the use of a heavy-flavor tagging algorithm to enhance the signal with respect to background.
The analysis
presented here updates the previously published result~\cite{2l2qpaper} by the use of the most recent theoretical predictions for the
Higgs boson mass lineshape and the correction of a problem in the background
description. The measurement in this channel
 uses the same $\sqrt{s}=7\TeV$ data set as the published paper~\cite{2l2qpaper} and uses
the same selection requirements.

Reconstructed electrons and muons are required to have $\PT > 40 (20)\GeV$ for the
highest-$\pt$ (second-highest-$\pt$)
lepton. Electrons (muons) are required to have $\abs{\eta} < 2.5 (2.4)$, with the transition
region between ECAL barrel and endcap, $1.44 < \abs{\eta} < 1.57$, excluded for electrons.
Jets are required to have
$\PT > 30\GeV$ and $\abs{\eta} < 2.4$. Each pair of oppositely-charged leptons of the same flavor,
and each pair of jets,
are considered as $\cPZ$ candidates. Background contributions are reduced by requiring $75 < \mjj < 105\GeV$ and $70 < \mll < 110\GeV$.

In order to exploit the different jet composition of signal and background, events are classified into three
mutually exclusive categories, according to the number of selected $\cPqb$-tagged jets: 0b-tag, 1b-tag and 2b-tag.
An angular likelihood discriminant is used to separate signal-like from background-like events in each category~\cite{Gao:2010qx}. A ``quark-gluon'' likelihood discriminant ($\cPq\Pg$LD), intended to distinguish gluon
jets from light-quark jets, is employed for the 0b-tag category, which is expected to be dominated by $\cPZ+\text{jets}$
background. A requirement on the qgLD value reduces backgrounds by approximately 40\% without any loss in
the signal efficiency. In order to suppress the substantial $\ttbar$ background in the 2b-tag category, a discriminant
$\lambda$ is used. This variable is defined as the ratio of the likelihoods of a hypothesis with $\MET$ equal to the value measured with
the PF algorithm, and the null hypothesis $\MET=0\GeV$~\cite{Chatrchyan:2011tn}. This discriminant provides a measure of
whether the event contains genuine missing transverse energy. Events in the 2b-tag category are required to
have $2\ln{\lambda} < 10$. When an event contains multiple $\cPZ$ candidates passing the selection requirements,
only the ones with jets in the highest b-tag category are retained for analysis. If multiple candidates are
still
present, the ones with $\mjj$ and $\mll$ values closest to the $\cPZ$ mass are retained.

The statistical analysis is based on the invariant mass of the Higgs boson candidate, $\mZZ$,
applying the constraint that the dijet invariant mass is consistent
with that of the $\Zo$ boson. Data containing a Higgs boson signal are expected to show a resonance peak
over a
continuum background distribution.

The background distributions are estimated from the  $\mjj$ sidebands, defined as $60<\mjj<75\GeV$ and
$105 <\mjj<130\GeV$. In simulation, the composition and distribution of the dominant backgrounds in the sidebands
are observed to be similar to those in the signal region. The distributions derived from data sidebands are measured
for each of the three b-tag categories and used to estimate the normalization of the background and its dependence on
$\mZZ$. The results of the sideband interpolation procedure are in good agreement with the observed distributions in
data. In all cases, the dominant backgrounds include $\cPZ+\text{jets}$ with either light- or heavy-flavor jets and \ttbar
background, both of which populate the $\mjj$ signal region and the $\mjj$ sidebands. The diboson background amounts
to less than 5\% of the total in the 0b and 1b-tag categories, and about 10\% in the 2b-tag category. No significant
difference is observed between results from data and the background expectation.
\begin{figure}[htbp]
\begin{center}
\includegraphics[width=\cmsFigWidth]{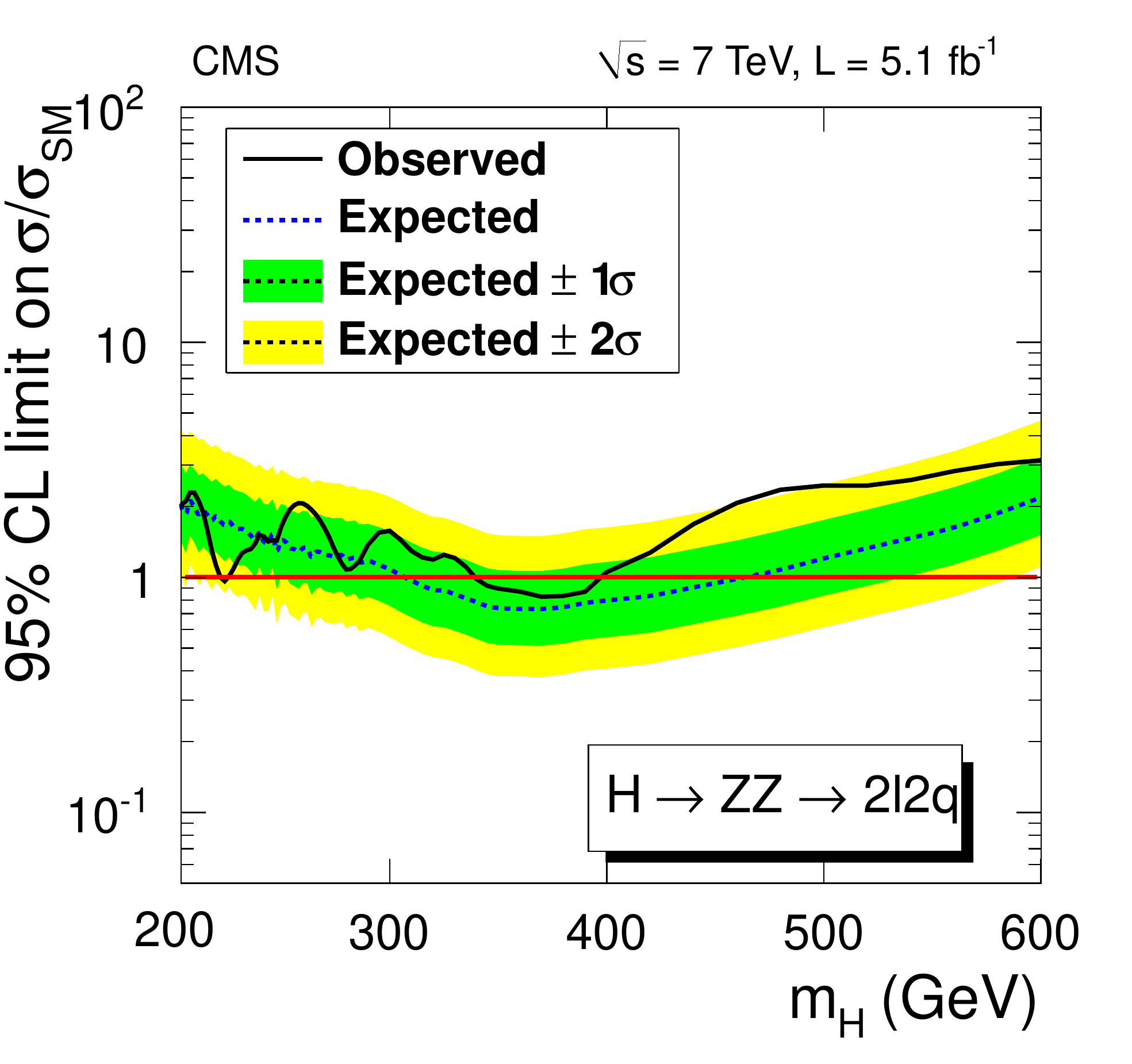}
\caption{
Observed (solid line) and expected (dashed line) 95\% CL upper limit on the ratio of the product of the production cross section
and branching fraction, to the SM expectation for the Higgs boson in the $\mathrm{H} \to \ZZ \to 2\ell2\cPq$ channel.
}
\label{fig:ZZ2l2qlimit}
\end{center}
\end{figure}

The distribution of $\mZZ$ for the background is parametrized by an empirical function
constructed of a Crystal Ball distribution~\cite{CrystalBall1,CrystalBall2,CrystalBall3}
multiplied by a Fermi function, $f(\mZZ) = 1/[1+\re^{-(\mZZ-a)/b}]$,
fitted to the shape and with normalization determined from the sidebands.
The dominant normalization uncertainty in the background estimation
is due to statistical uncertainty of the number of events in the sidebands.
The reconstructed signal distribution has two components.
The Double Crystal Ball function~\cite{CrystalBall1,CrystalBall2,CrystalBall3} is
used to describe the events with well reconstructed Higgs boson decay products.
The $\mZZ$ spectrum for misreconstructed events is described with a triangle function with linear rising and falling edges, convoluted with Crystal Ball function for better description of
 the peak and tail regions.
The signal reconstruction efficiency and the $\mZZ$ distribution are parametrized as a
function of $\mH$. The main uncertainties in the signal $\mZZ$ parametrization
are due to experimental resolution, which is predominantly due to the uncertainty on the jet energy scale~\cite{Chatrchyan:2011ds}. Uncertainties in b-tagging
efficiency are evaluated with a sample of jet events enriched in
heavy flavors by requiring a muon to be spatially close to a jet. The uncertainty associated with the $\cPq\Pg$LD
selection efficiency is evaluated using the $\Pgg+\text{jet}$ sample in data, which predominantly contains light quark jets.

The upper limits at 95\% CL on the ratio of the production cross section for the Higgs boson to the SM expectation,
obtained from the combination of all categories, are presented in Fig.~\ref{fig:ZZ2l2qlimit}.
This exclusion limit supersedes the previously published one~\cite{2l2qpaper}.

\subsection{\texorpdfstring{$\PH \to \ZZ \to 2\ell 2\nu$}{H to ZZ to 2 ell 2 nu}}

This analysis identifies Higgs boson decays to a pair of $\cPZ$ bosons, with one of $\cPZ$ bosons
decaying leptonically and the other to neutrinos.
A detailed description of the analysis can be found in~\cite{Chatrchyan:2012ft}.
The
analysis strategy is based on a set of $\mH$-dependent selection requirements
applied on  \MET and $m_{\mathrm{T}}$, where
\ifthenelse{\boolean{cms@external}}{
\begin{multline*}
\mt^2 = \left [ \sqrt{(\pt^{\ell\ell})^2+m_{\ell\ell}^2}+\sqrt{(\MET)^2+m_{\ell\ell}^2} \right ]^2 - \\\left [ \vecPtell+\vecEtm \right ]^2.
\end{multline*}

}{
\begin{equation*}
\mt^2 = \left [ \sqrt{(\pt^{\ell\ell})^2+m_{\ell\ell}^2}+\sqrt{(\MET)^2+m_{\ell\ell}^2} \right ]^2 - \left [ \vecPtell+\vecEtm \right ]^2.
\end{equation*}
}
Events are required to have a pair of well identified, isolated leptons of same flavor ($\Pep\Pem$ or $\Pgmp\Pgmm$),
each with $\PT > 20\GeV$, with an invariant mass within a $30\GeV$ window centered on the $\cPZ$ mass. The $\PT$
of the dilepton system is required to be greater than $55\GeV$. Jets are considered only if
they have $\pt>30\GeV$ and $\abs{\eta}<5$. The presence of large missing transverse energy in the
event is also an essential feature of the signal.

To suppress $\cPZ+\text{jets}$ background, events are excluded from the analysis if the angle in the azimuthal plane between
the \vecEtm and the closest jet is smaller than 0.5 radians. In order to remove events where the lepton is mismeasured, events are rejected if $\MET > 60\GeV$ and $\Delta\phi(\ell,\vecEtm) < 0.2$. The top-quark background is suppressed by applying a veto on events having a b-tagged jet with $\pt > 30\GeV$ and $\abs{\eta} < 2.4$.
To further suppress the top-quark background, a veto is applied on events containing a ``soft muon'', with $\PT > 3\GeV$, which is typically produced in the leptonic decay of a bottom quark. To reduce the $\PW\cPZ$ background, in which both bosons
decay leptonically, any event with a third lepton ($\Pe$ or $\Pgm$) with $\PT > 10\GeV$, and passing the identification and
isolation requirements, is rejected.

The search is carried out in two mutually exclusive categories.
The VBF category contains events with at least two jets
with $|\Delta\eta_{\mathrm{jj}}|>4$
and $\mjj>500\GeV$.
Both leptons forming the $\cPZ$ candidate are required to lie
in this $\Delta\eta_{\mathrm{jj}}$ region, and
there should be no other jets in it.
The gluon fusion category includes all events failing the VBF selection, and is subdivided into subsamples according to the presence or absence of
reconstructed jets. The event categories are chosen in order to optimize the expected cross section
limit. In the case of the VBF category, a constant $\MET>70\GeV$ and no $\mt$ requirement are used, as no gain in sensitivity is obtained with a $\mH$-dependent selection.

\begin{figure}[htbp]
\begin{center}
{\includegraphics[width=0.48\textwidth]{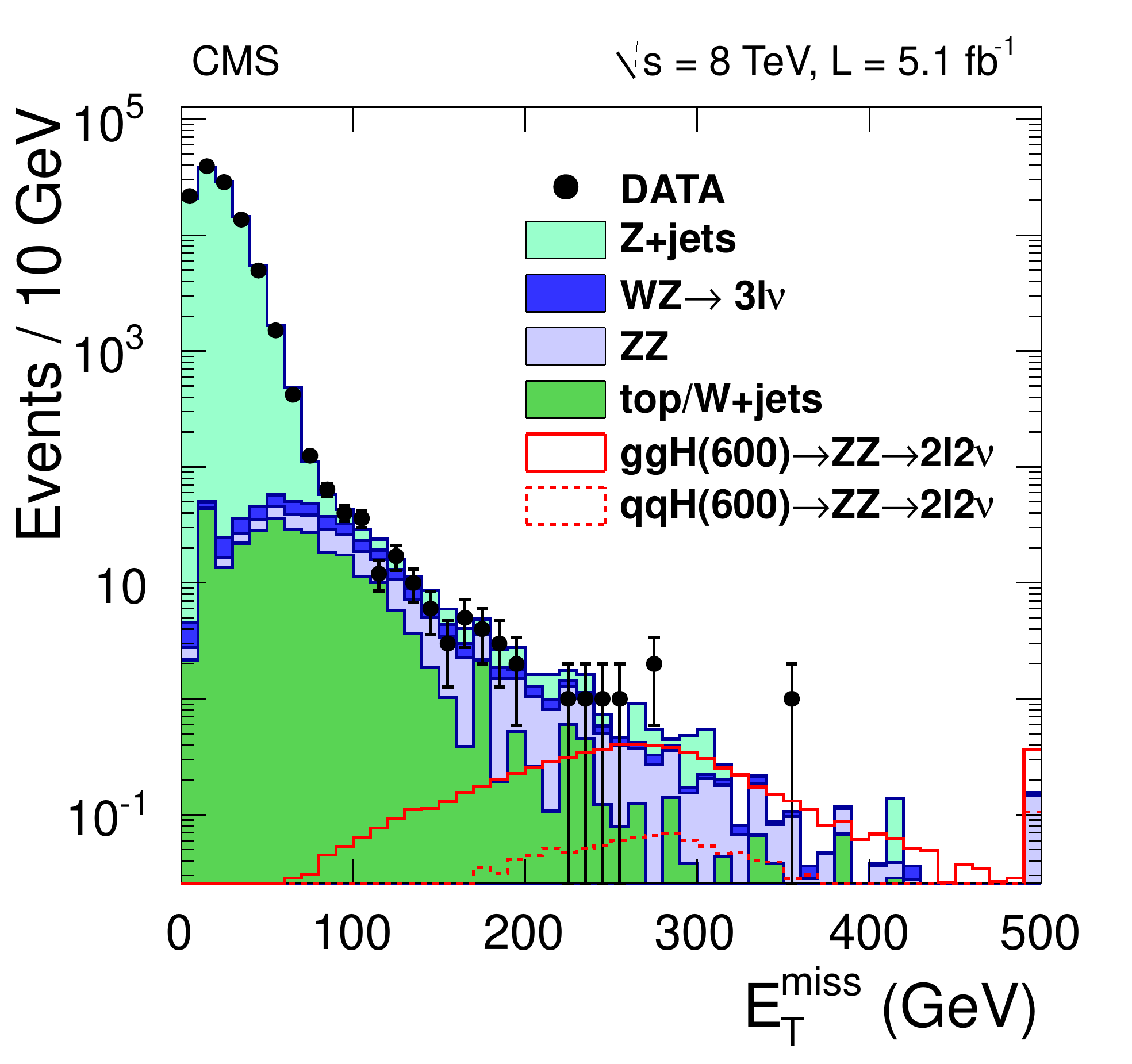}}
{\includegraphics[width=0.48\textwidth]{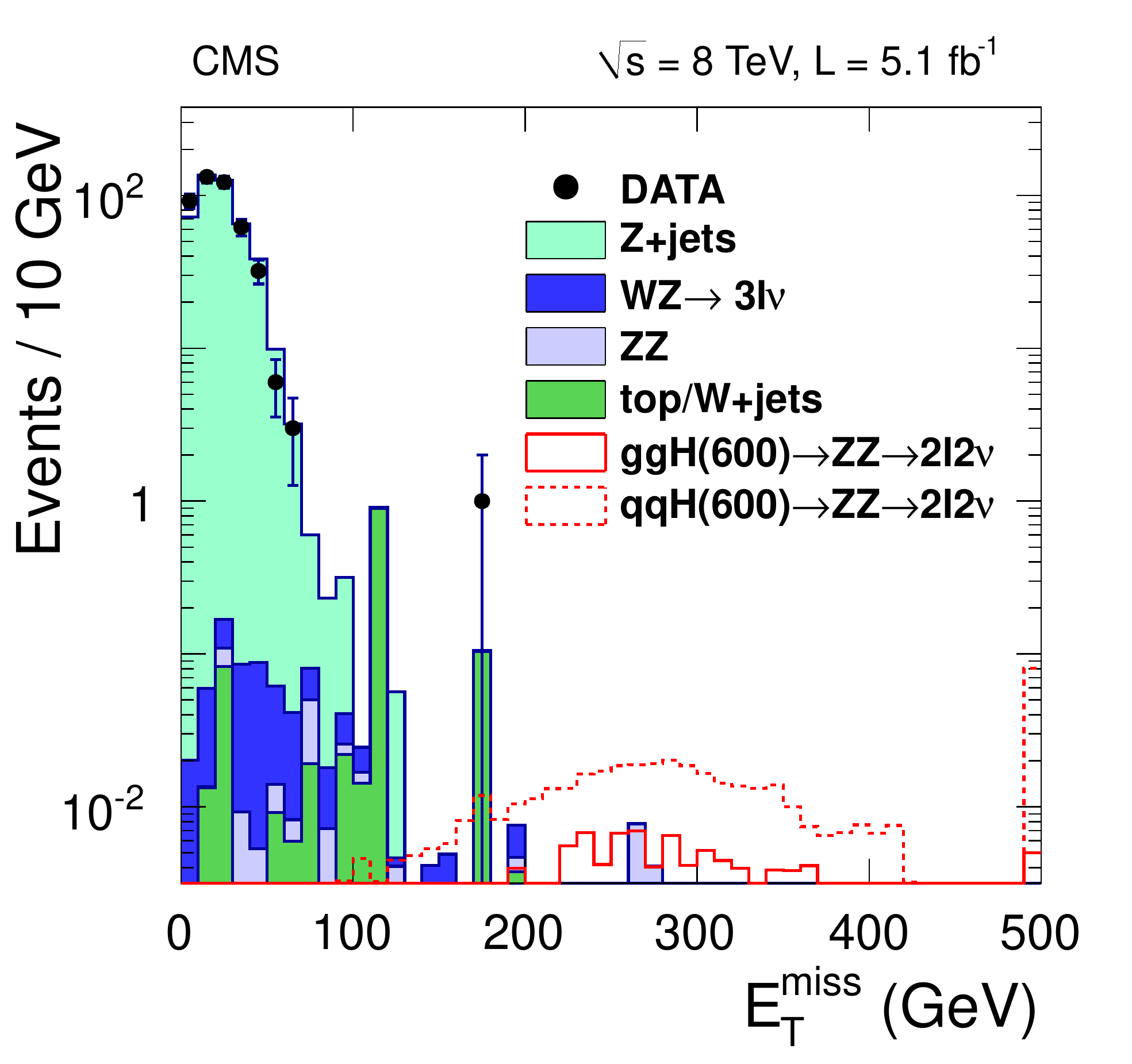}}
\caption{The \MET distribution in data compared to the estimated background
in the (\cmsLeft) gluon fusion and (\cmsRight) VBF categories {of the $\PH \to \ZZ \to 2\ell 2\nu$ channel}.
The dielectron and dimuon channels are combined.
Contributions from $\ZZ$, $\PW\cPZ$, non-resonant background and $\cPZ+\text{jets}$ background are stacked
on top of each other. The \MET distribution in signal events for $\mH = 600$\GeV is also shown. The last bin
in each plot contain the overflow entries.}
\label{fig:zgamma_met_data}
\end{center}
\end{figure}

The background composition is expected to vary with the hypothesised value of $\mH$. At low $\mH$, $\cPZ+\text{jets}$
and $\ttbar$ are the largest contributions, whilst at higher $\mH$ (above 400\GeV), the irreducible $\cPZ\cPZ$ and
$\PW\cPZ$ backgrounds dominate. The $\cPZ\cPZ$ and $\PW\cPZ$ backgrounds are taken from simulation~\cite{Sjostrand:2006za,Alwall:2007st} and are
normalized to their respective NLO cross sections. The $\cPZ+\text{jets}$ background is modeled from a control
sample of $\gamma+\text{jets}$ events. This procedure yields an accurate model of
the \MET distribution in $\cPZ+\text{jets}$ events, shown in Fig.~\ref{fig:zgamma_met_data}.

The uncertainty associated with the $\cPZ+\text{jets}$ background estimate is affected by any residual contamination in the
$\Pgg+\text{jets}$ control sample from processes involving a photon and genuine $\MET$.
This contamination could be as large as 50\% of the total $\cPZ+\text{jets}$ background.
It is not subtracted,
but assigned a 100\% uncertainty.

\begin{figure}[htbp]
\begin{center}
\includegraphics[width=\cmsFigWidth]{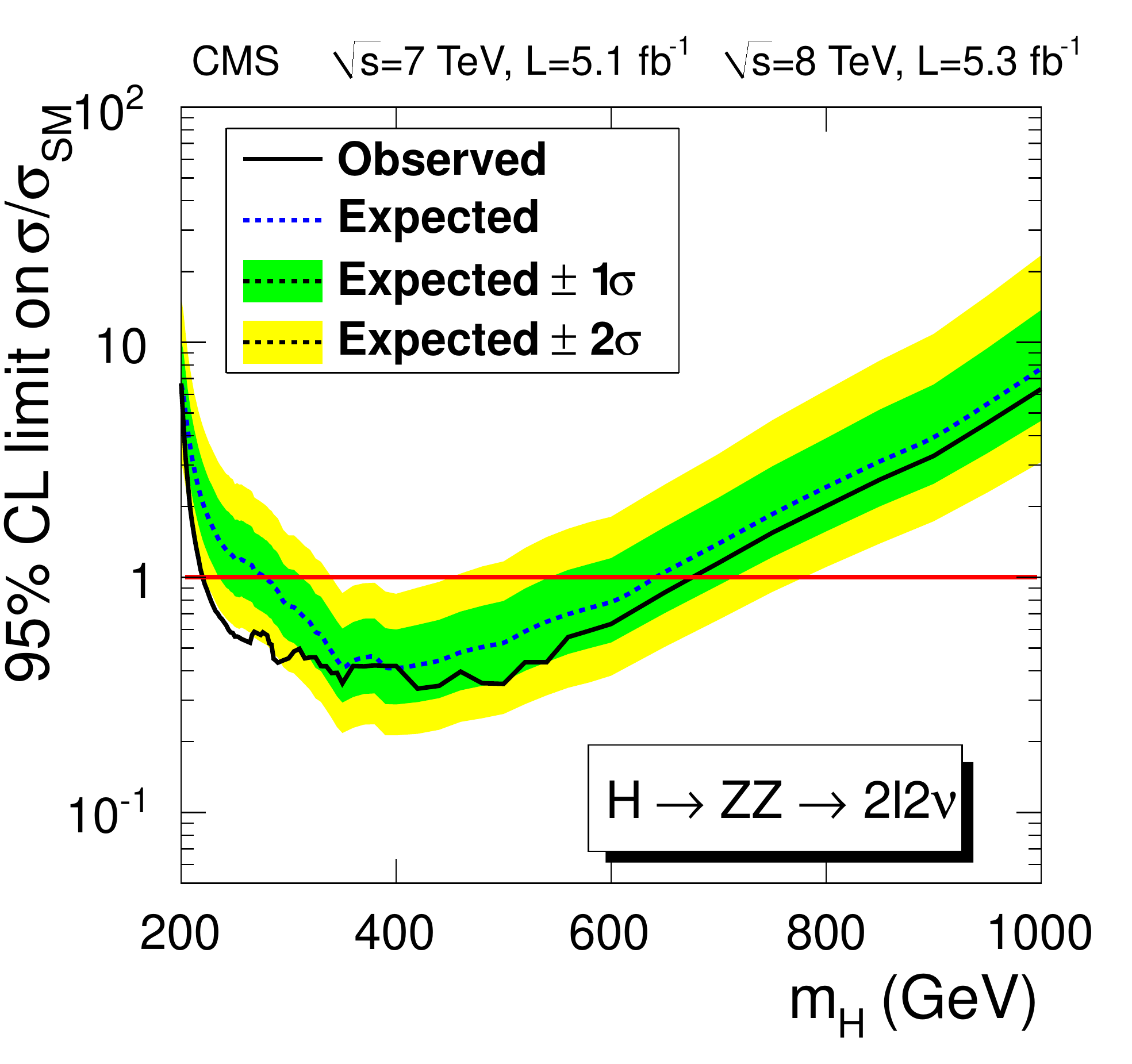}
\caption{Observed (solid line) and expected
(dashed line) 95\% CL upper limit on the ratio of the product of the  production cross
section and branching fraction to the SM expectation for the Higgs boson in the $\PH \to \ZZ \to 2\ell 2\nu$ channel.
}
\label{fig:limits_SM}
\end{center}
\end{figure}

Background processes that do not involve a $\cPZ$ resonance (non-resonant background) are estimated with a control
sample of events with dileptons of different flavor ($\Pe^{\pm}\Pgm^{\mp}$) that pass the full analysis selection.
This method cannot distinguish between the non-resonant background and a possible contribution from
$\PH \to \PW\PW \to 2\ell 2\cPgn$ events, which are treated as part of the non-resonant background
estimate. This treatment considers only the $\PH \to \cPZ\cPZ$ channel as signal and is combined
with the $\PH \to \PW\PW$ channel for the limit calculation.
The interference between $\cPZ\cPZ$ and $\PW\PW$ channels is also taken into
account~\cite{Chatrchyan:2012ft}. The non-resonant background in the $\Pep\Pem$ and $\Pgmp\Pgmm$ final states is estimated by applying a
scale factor to the selected $\Pe^{\pm}\Pgm^{\mp}$ events, estimated from the sidebands of the $\cPZ$ peak events
($40 < m_{\ell\ell} < 70\GeV$ and $110 < m_{\ell\ell} < 200\GeV$). The uncertainty associated with the estimate of the
non-resonant background is evaluated to be 25\%. 
No significant excess of events is observed over
the SM background expectation.
The observed and expected upper limits as a function of $\mH$ are shown in Fig.~\ref{fig:limits_SM}.

\section{Combined results}
\label{sec:results}

\begin{figure}[htbp`]
  \centering
\includegraphics[width=0.48\textwidth]{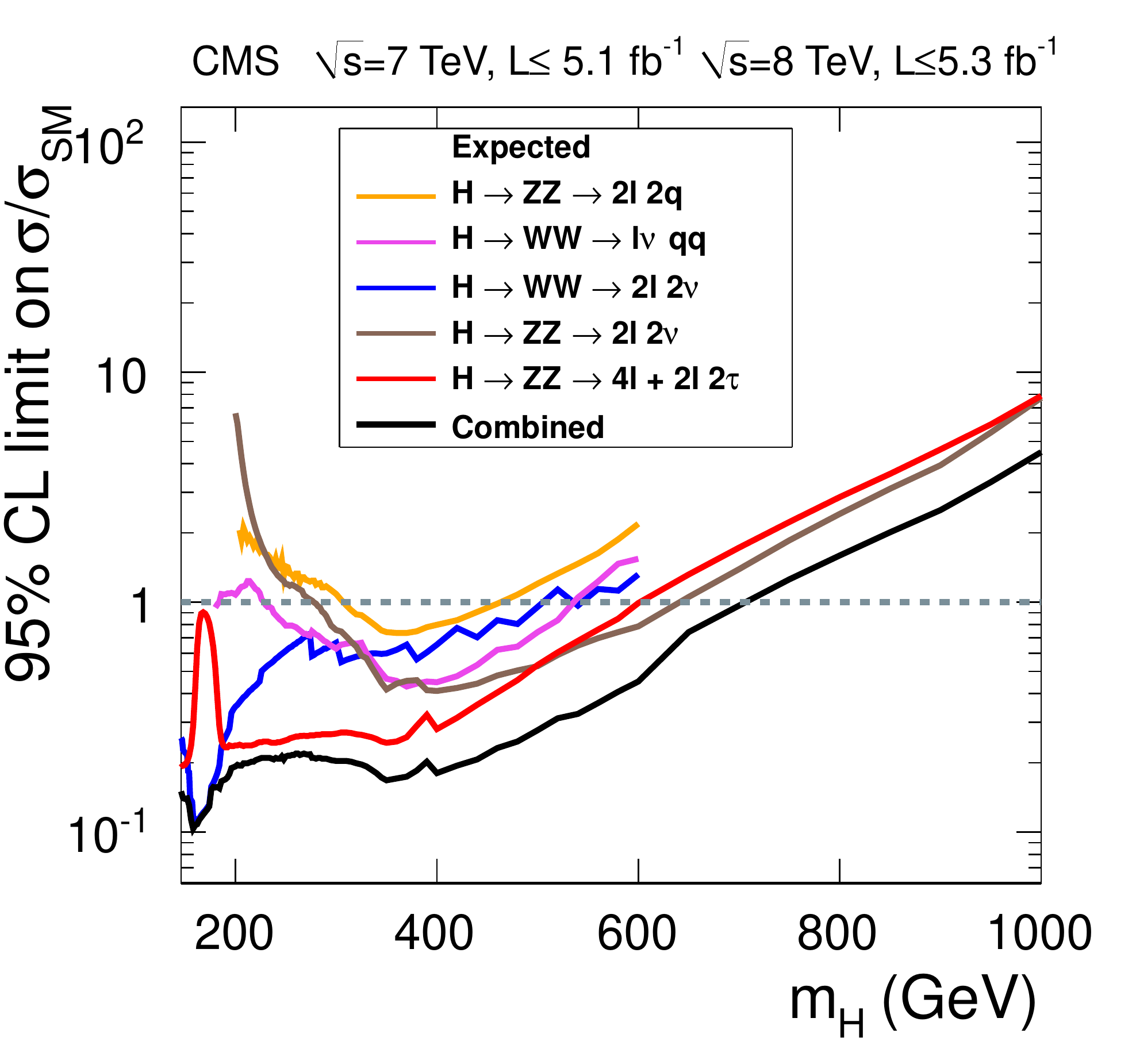}
\includegraphics[width=0.48\textwidth]{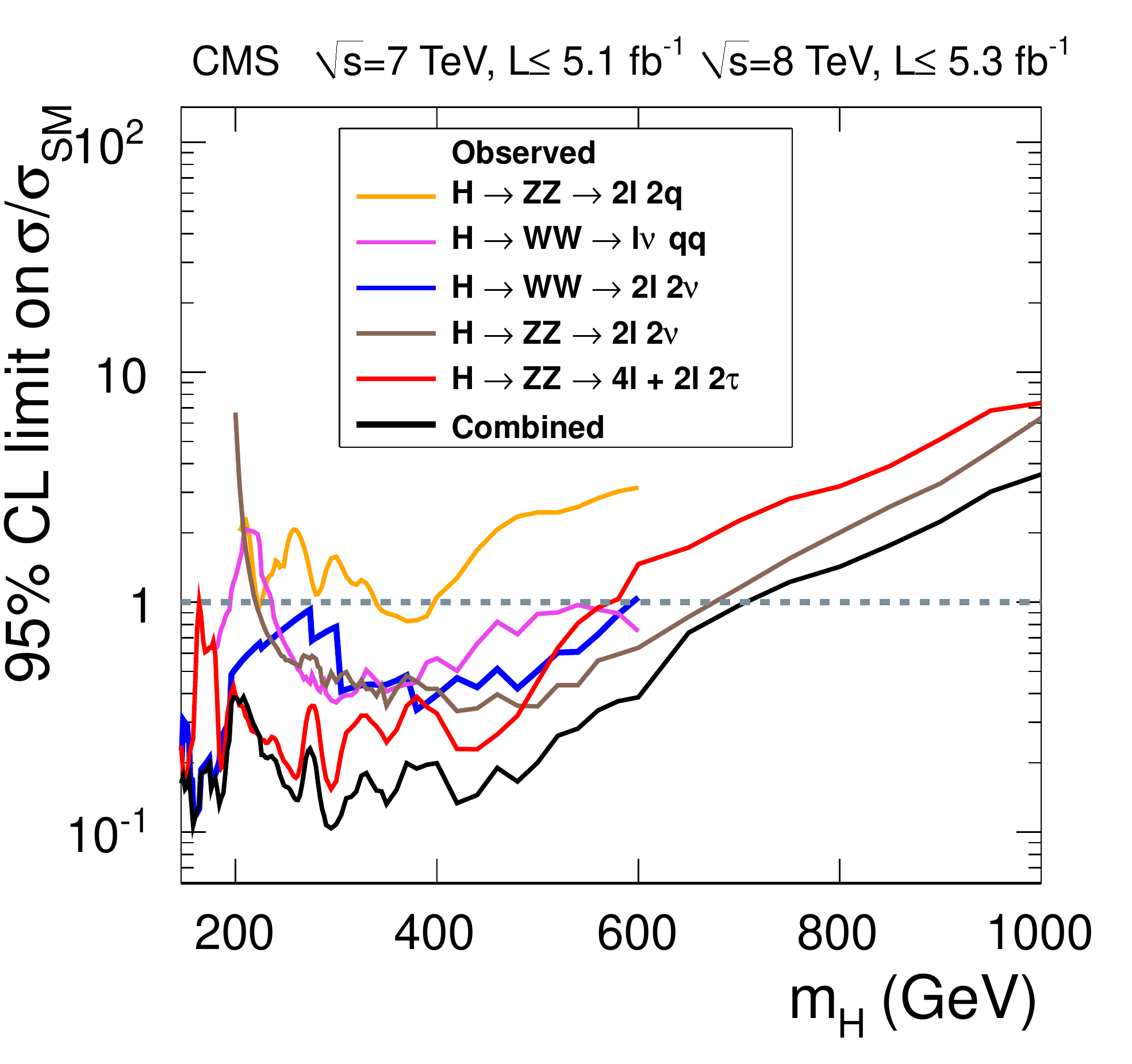}
  \caption{\label{fig:AllLimits} (\cmsLeft) Expected and  (\cmsRight) observed 95\% CL limits
  for all individual channels and their combination. The horizontal dashed line at unity
    indicates the SM expectation.
}
\end{figure}

\begin{figure}[htbp]
  \centering
  \includegraphics[width=\cmsFigWidth]{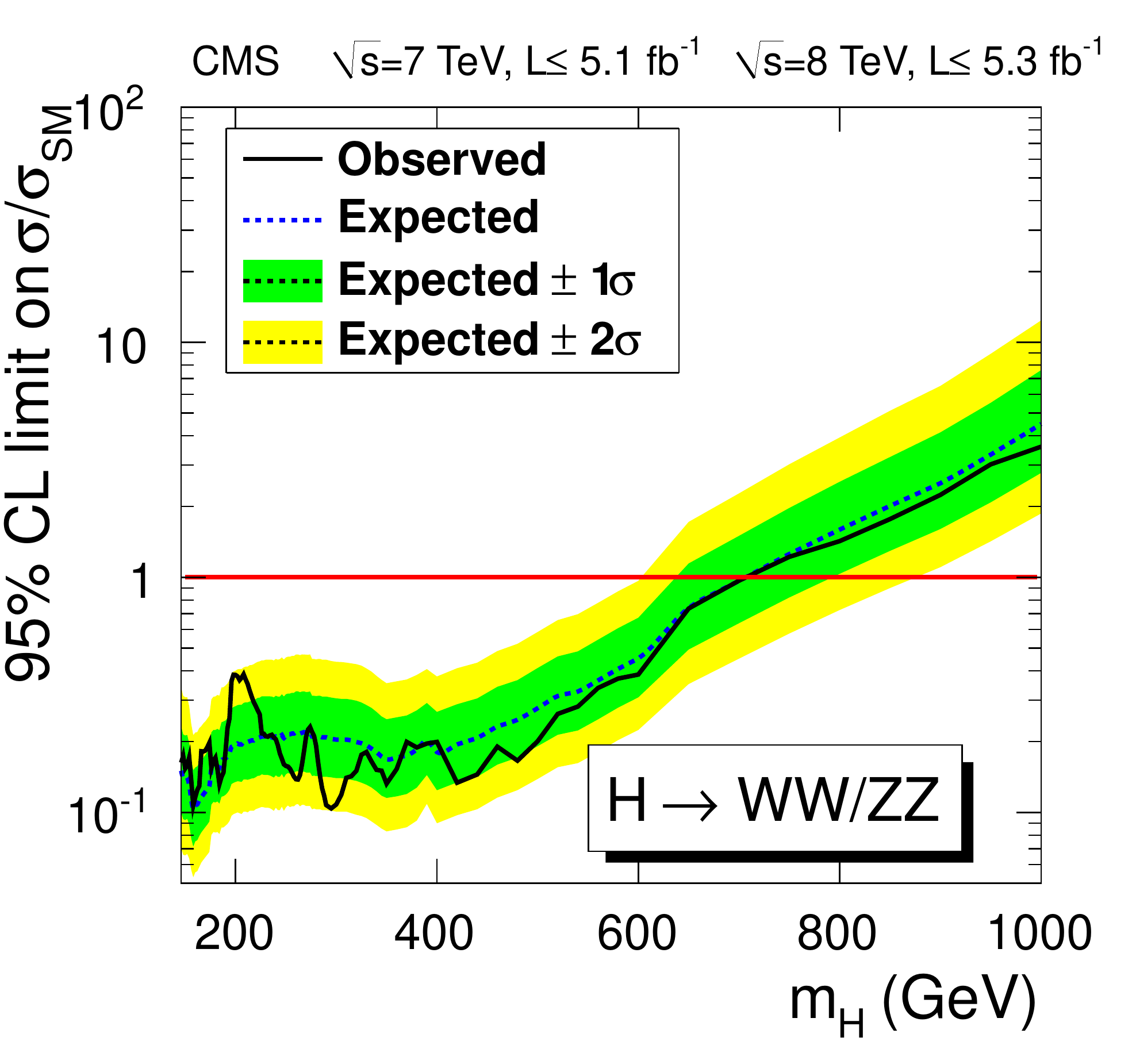}
  \caption{\label{fig:combined}Observed (solid line) and expected
  (dashed line) 95\% CL upper limit on the ratio of the production cross
  section to the SM expectation for the Higgs boson with all $\WW$ and $\ZZ$
  channels combined.}
\end{figure}

The expected and observed upper limits on the ratio of the production cross section for the Higgs boson to the SM expectation, for each of the individual channels presented in this paper, are shown in Fig.~\ref{fig:AllLimits}.
This figure also shows a combined limit, calculated using the methods outlined in Ref.~\cite{LHC-HCG-Report, Chatrchyan:2012tx}.
The combination procedure assumes the relative branching fractions to be those predicted by the SM, and takes into account the statistical and experimental systematic uncertainties as well as theoretical uncertainties.
In the mass region $145<\mH<200\GeV$ the branching fraction of the most sensitive channel, $\PH\to\cPZ\cPZ$, is decreasing and has a typical
dependence on $\mH$, which is reflected in both the expected and observed limits. In this mass region
the result of the combination is determined by the $\PW\PW\to\ell\nu\ell\nu$ channel.
At masses above 200\GeV the $\cPZ\cPZ\to 2\ell 2\ell'$ channel becomes dominant,
since
low background contributions in this channel allow to keep high
efficiency of the selection requirements.
Starting at approximately 400\GeV the $\cPZ\cPZ\to 2\ell 2\nu$ starts to contribute significantly. 
The branching fraction of $\cPZ\cPZ\to 2\ell2\nu$ is higher than
$\cPZ\cPZ\to 2\ell 2\ell'$, and the major
background contributions decrease with $\mH$ increase, thus allowing for selection requirements to
be more and more effective in the $2\ell 2\nu$ channel.
The combined observed and expected limits agree well within uncertainties
 as shown in Fig.~\ref{fig:combined}.

The previously expected exclusion range at 95\% CL, 118--543\GeV, is extended up to 700\GeV.
Previously published results exclude at 95\% CL the SM-like Higgs boson in the range $127 < \mH <  600\GeV$~\cite{Chatrchyan:2012tx}.
The results of this analysis extend the upper exclusion limit to $\mH = 710\GeV$.

\section{Summary}
\label{sec:summary}

Results are presented from searches for a standard-model-like Higgs boson in $\PH \to \WW$ and $\PH \to \ZZ$
decay channels, for Higgs boson mass hypotheses in the range $145 < \mH < 1000$\GeV. The analysis uses proton-proton
collision data recorded by the CMS detector at the LHC, corresponding to integrated luminosities of up to
 5.1\fbinv at
$\sqrt{s} = 7\TeV$ and up to 5.3\fbinv at $\sqrt{s} = 8\TeV$. The final states analysed include
two leptons and two neutrinos, $\PH \to \WW \to \ell\nu\ell\nu$ and $\PH \to \ZZ \to 2\ell 2\nu$, a lepton,
a neutrino, and two jets,
$\PH \to \WW \to \ell \nu \cPq\cPq$, two leptons and two jets, $\PH \to \ZZ \to 2\ell 2\cPq$,
and four leptons, $\PH \to \ZZ \to 2\ell 2\ell'$, where $\ell = \Pe$ or $\Pgm$ and $\ell' = \Pe$ or $\Pgm$, or $\Pgt$.
The results  are consistent with standard model background expectations.
The combined
upper limits at 95\% confidence level on products of the cross section and branching
fractions
exclude a standard-model-like Higgs boson in the range
$145 < \mH <  710\GeV$, thus extending the
mass region excluded by CMS from
127--600\GeV up to 710\GeV.

\section*{Acknowledgements}

{\tolerance=600 We congratulate our colleagues in the CERN accelerator departments for the excellent performance of the LHC and thank the technical and administrative staffs at CERN and at other CMS institutes for their contributions to the success of the CMS effort. In addition, we gratefully acknowledge the computing centres and personnel of the Worldwide LHC Computing Grid for delivering so effectively the computing infrastructure essential to our analyses. Finally, we acknowledge the enduring support for the construction and operation of the LHC and the CMS detector provided by the following funding agencies: BMWF and FWF (Austria); FNRS and FWO (Belgium); CNPq, CAPES, FAPERJ, and FAPESP (Brazil); MEYS (Bulgaria); CERN; CAS, MoST, and NSFC (China); COLCIENCIAS (Colombia); MSES (Croatia); RPF (Cyprus); MoER, SF0690030s09 and ERDF (Estonia); Academy of Finland, MEC, and HIP (Finland); CEA and CNRS/IN2P3 (France); BMBF, DFG, and HGF (Germany); GSRT (Greece); OTKA and NKTH (Hungary); DAE and DST (India); IPM (Iran); SFI (Ireland); INFN (Italy); NRF and WCU (Republic of Korea); LAS (Lithuania); CINVESTAV, CONACYT, SEP, and UASLP-FAI (Mexico); MSI (New Zealand); PAEC (Pakistan); MSHE and NSC (Poland); FCT (Portugal); JINR (Armenia, Belarus, Georgia, Ukraine, Uzbekistan); MON, RosAtom, RAS and RFBR (Russia); MSTD (Serbia); SEIDI and CPAN (Spain); Swiss Funding Agencies (Switzerland); NSC (Taipei); ThEPCenter, IPST and NSTDA (Thailand); TUBITAK and TAEK (Turkey); NASU (Ukraine); STFC (United Kingdom); DOE and NSF (USA).

Individuals have received support from the Marie-Curie programme and the European Research Council and EPLANET (European Union); the Leventis Foundation; the A. P. Sloan Foundation; the Alexander von Humboldt Foundation; the Belgian Federal Science Policy Office; the Fonds pour la Formation \`a la Recherche dans l'Industrie et dans l'Agriculture (FRIA-Belgium); the Agentschap voor Innovatie door Wetenschap en Technologie (IWT-Belgium); the Ministry of Education, Youth and Sports (MEYS) of Czech Republic; the Council of Science and Industrial Research, India; the Compagnia di San Paolo (Torino); and the HOMING PLUS programme of Foundation for Polish Science, cofinanced from European Union, Regional Development Fund.\par}

\bibliography{auto_generated}   

\cleardoublepage \appendix\section{The CMS Collaboration \label{app:collab}}\begin{sloppypar}\hyphenpenalty=5000\widowpenalty=500\clubpenalty=5000\textbf{Yerevan Physics Institute,  Yerevan,  Armenia}\\*[0pt]
S.~Chatrchyan, V.~Khachatryan, A.M.~Sirunyan, A.~Tumasyan
\vskip\cmsinstskip
\textbf{Institut f\"{u}r Hochenergiephysik der OeAW,  Wien,  Austria}\\*[0pt]
W.~Adam, T.~Bergauer, M.~Dragicevic, J.~Er\"{o}, C.~Fabjan\cmsAuthorMark{1}, M.~Friedl, R.~Fr\"{u}hwirth\cmsAuthorMark{1}, V.M.~Ghete, N.~H\"{o}rmann, J.~Hrubec, M.~Jeitler\cmsAuthorMark{1}, W.~Kiesenhofer, V.~Kn\"{u}nz, M.~Krammer\cmsAuthorMark{1}, I.~Kr\"{a}tschmer, D.~Liko, I.~Mikulec, D.~Rabady\cmsAuthorMark{2}, B.~Rahbaran, C.~Rohringer, H.~Rohringer, R.~Sch\"{o}fbeck, J.~Strauss, A.~Taurok, W.~Treberer-treberspurg, W.~Waltenberger, C.-E.~Wulz\cmsAuthorMark{1}
% ...
\vskip\cmsinstskip
% ...
1:~~Also at Vienna University of Technology, Vienna, Austria\\
2:~~Also at CERN, European Organization for Nuclear Research, Geneva, Switzerland\\
% ...

\end{sloppypar}
\end{document}